\documentclass[preprint]{aastex631}

\newcommand{\ALpaper}{AL21}

\def\revise#1{{ #1}}

\shortauthors{Ashfield et al.}

\graphicspath{{./}{figures/}}

\begin{document}

\title{Connecting Chromospheric Condensation Signatures to Reconnection Driven Heating Rates in an Observed Flare}

\accepted{December 3rd, 2021}
\submitjournal{ApJ}

\author{William H. Ashfield IV}
\affiliation{Dept.\ of Physics, Montana State University,
Bozeman, MT 59717}

\author{Dana W. Longcope}
\affiliation{Dept.\ of Physics, Montana State University,
Bozeman, MT 59717}

\author{Chunming Zhu}
\affiliation{Dept.\ of Physics, Montana State University,
Bozeman, MT 59717}

\author{Jiong Qiu}
\affiliation{Dept.\ of Physics, Montana State University,
Bozeman, MT 59717}

%==========================================================
%==========================================================

\begin{abstract}

Observations of solar flare reconnection at very high spatial and temporal resolution can be made indirectly at the footpoints of reconnected loops into which flare energy is deposited.  The response of the lower atmosphere to this energy input includes a downward-propagating shock called chromospheric condensation, which can be observed in the UV and visible.  In order to characterize reconnection using high-resolution observations of this response, one must develop a quantitative relationship between the two. Such a relation was recently developed and here we test it on observations of chromospheric condensation in a single footpoint from a flare ribbon of the X1.0 flare on 25 Oct.\ 2014 (SOL2014-10-25T16:56:36).  Measurements taken of Si\,\textsc{iv}\,1402.77\,\AA\ emission spectra using the Interface Region Imaging Spectrograph (IRIS) in a single pixel show red-shifted component undergoing characteristic condensation evolution. We apply the technique called the Ultraviolet Footpoint Calorimeter (UFC) to infer energy deposition into the one footpoint. This energy profile, persisting much longer than the observed condensation, is input into a one-dimensional, hydrodynamic simulation to compute the chromospheric response, which contains a very brief condensation episode. From this simulation we synthesize Si\,\textsc{iv} spectra and compute the time-evolving Doppler velocity. The synthetic velocity evolution is found to compare reasonably well with the IRIS observation, thus corroborating our reconnection-condensation relationship. The exercise reveals that the chromospheric condensation characterizes a particular portion of the reconnection energy release rather than its entirety, and that the time scale of condensation does not necessarily reflect the time scale of energy input. 

\end{abstract}

%==========================================================
%==========================================================
%==========================================================
%==========================================================
\section{Introduction} \label{sec:intro}

A solar flare is a sudden increase in broad spectrum radiation arising from the release of magnetic free energy in the corona. Magnetic reconnection, although yet to be directly observed due to the difficulty of coronal magnetic field measurement, produces many indirect signatures from which information about the flare energy release process might be inferred \citep{fletcher2011}. One such signature is the rapid enhancement of UV emission at the flare ribbon formed from the footpoints of newly formed flare loops.  Enhancement occurs as the result of energy transported from the reconnection site to the lower solar atmosphere --- either by Alfvén waves \citep{emslie1982,fletcher2008}, non-thermal particles \citep{syrovatskii1972}, or thermal conduction \citep{craig1976,antiochos1978}. This energy deposited in lower atmospheric layers heats and ablates large amounts of dense material upward to fill coronal loops in a process known as chromospheric evaporation \citep{canfield1980,antonucci1982}. As evaporation leads to enhanced coronal soft X-ray and EUV emissions, the energy released via reconnection can be subsequently mapped to the coronal flare response \citep{neupert1968,qiu2021}.

Corollary to evaporation are the plasma downflows observed in the lower atmosphere, commonly referred to as chromospheric condensation \citep{acton1982,fisher1985_1}. This phenomenon, seen as another direct consequence of energy deposition, has proven to be a useful diagnostic for inferring the properties flare energy release given its amenability to observation at high cadence and spatial resolution in UV and optical wavelengths (see review by \cite{depontieu2021}). The first diagnostic investigations used observations of redshifted asymmetries in H$\alpha$ spectral lines to infer the value of coronal energy flux reaching the chromosphere \citep{zarro1989,canfield1990}. These studies were contemporary with early numerical experiments that found a power-law relationship between the condensation peak downflow velocity and the energy flux deposited in the corona \citep[e.g.][]{fisher1987,fisher1989}. More recently, \citet{longcope2014} used a strictly thermal model to find a similar scaling law for the peak downflow velocity, which was further confirmed for fluxes $\lesssim10^{10}$\,erg\,cm$^{-2}$\,s$^{-1}$ in stratified atmospheres by \citet[hereafter \ALpaper]{ashfield2021}.

In common practice, the diagnostic application of condensation investigations typically involves comparisons between observed spectral lines and those synthesized from the results of numerical flare models. Many of these models focus only on energy transport via non-thermal electrons. In a data-driven approach, energy flux arising from electron precipitation in the chromosphere is first inferred from a combination of spatially resolved hard X-ray (HXR) and UV spectra, which is then used to drive hydrodynamic flare loop simulations \citep{rubio2015,rubio2015_2,kuridze2015,kowalski2017}. By reproducing observed line profiles using the modeled hydrodynamic response to injected electrons, the properties of flare energy release as they pertain to non-thermal electron beams, and thus the mechanism for particle acceleration during reconnection, can be further constrained. 

In addition to the energy flux delivered to the chromosphere, the time scale on which reconnection and flare energy release occurs remains an unknown property that one might hope to constrain using condensation observation. Many  theoretical models of reconnection conclude that magnetic energy conversion on a given flux tube occurs on time scales on the order of the Alfvén time across that tube (i.e.\ seconds, \cite{kopp1976,priest2002}). Consequently, many numerical flare models use this timescale for the duration of energy flux released. Studies focusing on electron precipitation have typically used \textsl{ad hoc} heating profiles with short duration beam pulses. However, the heating durations vary between studies, from the several seconds to over a minute \citep{nagai1984,fisher1985_2,reep2016,kerr2019s}. Recent data-driven investigations have likewise inferred a beam heating duration of $\sim20$\,s from HXR lightcurves \citep{rubio2016,graham2020}. In contrast, \cite{reep2018} found beam durations of $\sim50-200$\,s better reproduce redshifts seen in observations of prolonged condensation \citep{warren2016,li2017}.

Studies outside the scope of non-thermal electron deposition have also generated a range of flare energy release models. Early simulations with conduction driven condensation used \emph{ad hoc} heating functions whose time profiles were unique, such as the piecewise function in \citet{gan1991}, and lasted on the order of minutes \citep{cheng1983,macneice1986}. Other investigations have studied the retraction of a post-reconnection flare loop as a model for flare energy release \citep{longcope2009,guidoni2010}. The shortening of magnetic field lines cause a compression of plasma along the flare loop, giving rise to gas-dynamic shocks that generate a heat flux into the chromosphere. Because the tube retracts at Alfvénic speeds, the rate of energy release is on the order of seconds \citep{longcope2015}.

The variation found in heating profiles used by numerical investigations make it clear the connection between the time scale of flare energy release and the dynamics of condensation is not yet well understood. Furthermore, despite the advancements of diagnostic studies in constraining the energy flux intensity streaming from the corona, they primarily focus on energy delivered via non-thermal electrons. Although observational signatures and numerical experiments provide substantial evidence for this type of energy transport, there is evidence that it does not constitute the entirety of flare energy deposition in the chromosphere. One indication of this are flares with chromospheric flows occurring in the absence of co-spatial HXR emission  \citep{zarro1988,li2017}, or where flows form well after footpoint HXR emission has subsided \citep{czaykowska2001,battaglia2009}. In addition, observations have found impulsive rises in optical and UV emissions at footpoints prior to SXR emission and without HXR signatures \citep{warren2001,coyner2009}, suggesting evaporation, and consequently condensation, driven by means other than non-thermal electrons. The continued enhancement in SXR emission after HXR emissions have abated also suggest further heating of the corona in other forms is required \citep{veronig2002}. Because reconnection theory is contingent upon the behavior of these different signatures, it is paramount for condensation models to coincide with such observational signatures if they are to exist within the broader view of flare energy release.

Motivated by the key role of footpoint diagnostics, \cite{qiu2012} and \cite{liu2013} developed a practical technique for inferring flare energy release using spatially and temporally resolved UV emission at footpoints to derive heating profiles of individual flare loops. The UFC method, as the technique is called, further takes into account the coronal response to impulsive heating by modeling the evolution of flare loops and comparing synthesized SXR and EUV emissions to the observed lightcurves. The tracking of coronal emission in different wavelengths, namely the six coronal EUV bands from the Atmosphere Imaging Assembly (AIA) \citep{lemen2012} aboard the Solar Dynamic Observatory (SDO) and the two SXR bands of GOES, allows for a characterization of the plasma temperature evolution over time, thus connecting long term flare evolution to the initial energy release. Although studies utilizing this method have accurately reproduced coronal observations \citep{zhu2018,qiu2021}, chromospheric response to the inferred energy deposition has not been investigated outside the brief analysis presented in \cite{qiu2016}.  In fact, the UFC uses an empirical relation between the coronal energy input and the ensuing chromospheric response.  The success of this empirical approach strongly suggests a physical mechanism at work, but it has not yet been explored.

In the present work, we take up the challenge of demonstrating the potential of chromospheric condensation as a diagnostic of coronal energy release.  We do this by applying the theoretical results of \ALpaper\ to observations of condensation in a particular flare.  We use high-cadence, sit-and-stare observations made by IRIS \revise{\citep{depontieu2014}} of an X1.0 class flare.   These observation revealed asymmetric Si \textsc{iv} 1402.77\,\AA\ line profiles within a single pixel, with redshifts exceeding 40\,km\,s$^{-1}$ decaying on the order of tens of seconds.  Application of the \ALpaper\ to these observations allowed us to infer the energy flux generated in the corona.  

The object of the present work is to use numerical solutions to independently confirm that such an energy flux would in fact produce the observed condensation.  While we do succeed, we also uncover a potential limitation.  Chromospheric condensation occurs on its own short time scale independent of the time scale of energy release.  The time of the condensation, $\sim15$\,s, therefore gives no insight into the time scale of energy release.  When the latter occurs over a longer time, as we infer it to in this case, the chromospheric response can diagnose only a particular instant within the energy release and not its entirety.  We thereby confirm the accuracy of this diagnostic, but also reveal its limitation.

Our numerical simulation requires, as input, an energy flux profile and a mechanism by which this flux is carried.  We obtain both independently in order to produce a synthetic condensation event we can compare with observation.  We use RHESSI observations to characterize the nature of the energy flux.   In this data we find no indication of HXR emission emanating from the flare ribbon where the condensation is observed.  We take this as evidence that the energy flux driving the condensation is carried primarily by thermal conduction.  We quantify that energy flux, and its time profile, using the UFC model driven by observations from SDO/AIA and GOES.
Because UFC does not depend on non-thermal models of energy transport, its use is a natural extension for determining energy release in the event of condensation driven by thermal conduction. Additionally, the UFC-derived heating profile, when used to drive a hydrodynamic simulation of a flare loop, allows us to model condensation dynamics using the observed coronal flare response, given the AIA EUV lightcurves required as an input for UFC. To our knowledge, this is the first investigation \revise{to model energy deposition into the chromosphere using a heating profile derived from coronal emission measurements}.

In order to analyze energy release properties from observed condensation characteristics, we first identified the condensation event amongst the evolution  of the flare ribbon along the IRIS slit. The process of finding and analysing the dynamics of the downflows seen in Si \textsc{iv} line profiles is described in Section \ref{sec:cond}. The energy subjected to the single IRIS pixel containing the condensation event is inferred using the UFC method in Section \ref{sec:energy}. In Section \ref{sec:1D}, a one-dimensional hydrodynamic simulation is run using the inferred energy release as an input, with which we synthesize Si \textsc{iv} lines to compare to IRIS observations. The implications of our results are discussed in the final section. We find for the first time a gradual heating profile released in the corona produces a fast, impulsive response in the chromosphere.

\section{Condensation Observation} \label{sec:cond}

The single condensation event studied in this work took place during the 2014 October 25th X1.0 class flare in active region AR 12192.  This active region was apart of the largest sunspot region of solar cycle 24, producing six X-class flares during its duration. Due to the strength of these flares and AR 12192 having no observed coronal mass ejections, these so called \textsl{confined} flares have been the subject of previous investigations \citep{chen2015,inoue2016,li2019_cc}.

\begin{figure*}[ht!]
\gridline{\fig{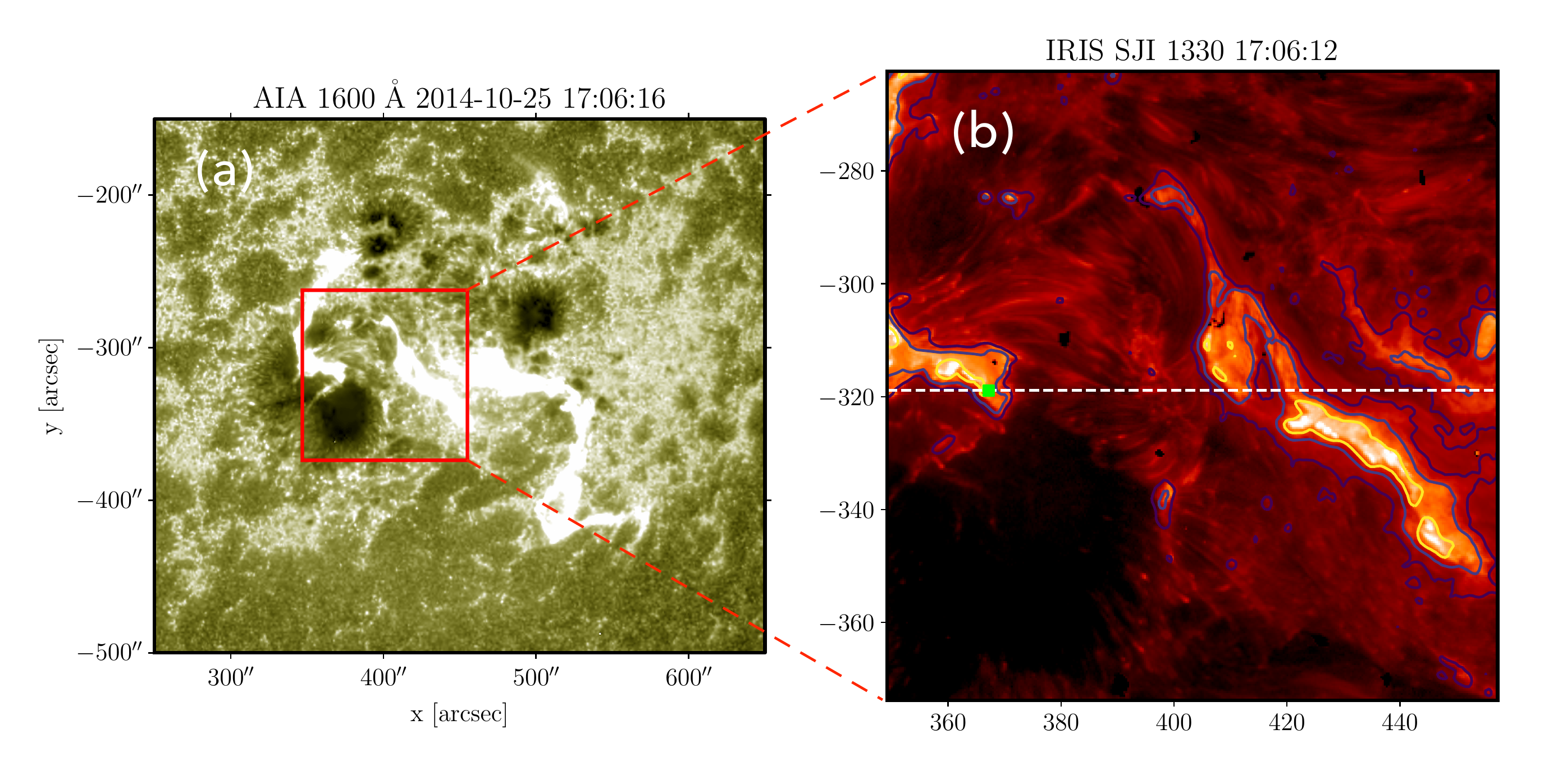}{\textwidth}{}}
\gridline{\fig{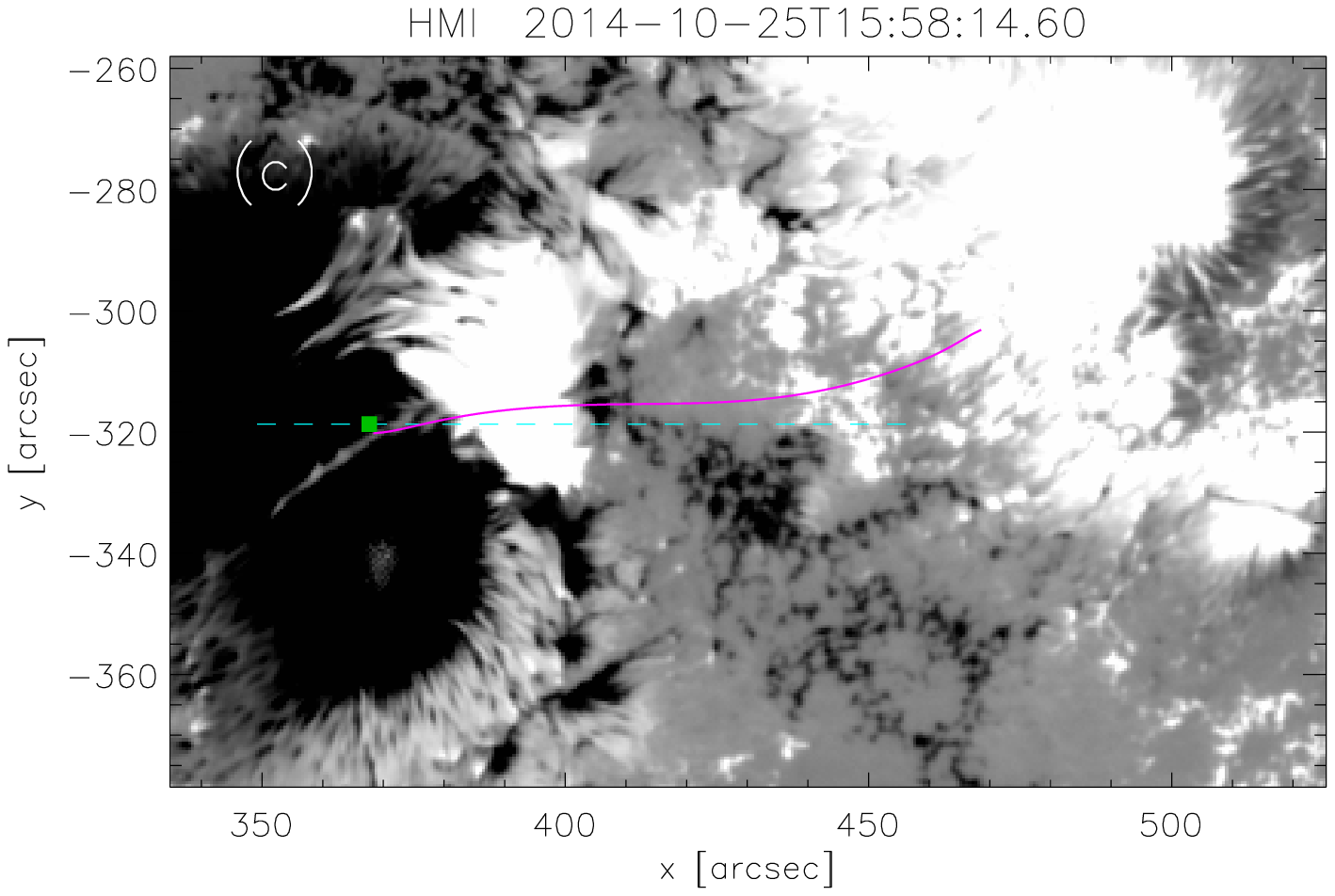}{\textwidth}{}}
\caption{(a) AIA\,1600\AA\,image showing the three flare ribbons during peak emission. IRIS SJI FOV is outlined in red, with the closest corresponding SJI 1330\AA\,image shown on the right (b). The IRIS slit is illustrated by the white dashed line and the selected pixel is given by the green square. Contours show [0.03, 0.08, 0.3] of maximum AIA\,1600\AA\,emission at 17:06:16. (c) HMI line-of-sight magnetogram. The magenta curve illustrates the field line produced in the extrapolation used to approximate the flare loop. The IRIS slit (cyan dashed line) and condensation pixel (green marker) are over plotted.
\label{fig:irisaia}}
\end{figure*}

The X1 flare SOL2014-10-25T17:08:00 was the fourth of these flares and was observed by SDO/AIA, as well as IRIS (OBSID 3880106953). The flare's east (negative\revise{-polarity}) ribbon is broken in two, giving the appearance of a flare with three ribbons. Figure \ref{fig:irisaia}a shows all three ribbons as brightenings in the AIA UV 1600\,\AA\ image during the period of peak emission. Figure \ref{fig:irisaia}b shows the temporally closest IRIS SJI image in 1330\,\AA\ with the position of the spectrograph slit shown along with \revise{coaligned} AIA 1600\,\AA\ contours. Here the IRIS spacecraft has been rolled such that the slit is in the E-W orientation. The slit was located over the western-most ribbon before and during the impulsive phase of the flare, \revise{but crossed a relatively weak segment ($\sim30$\% of maximum 1600\,\AA\ ribbon emission), providing data without much saturation. Running in sit-and-stare mode,} the IRIS spectrograph data contained the full flare line list with the high cadence of 5\,s with 4\,s exposure time. This data included the \revise{FUV} 1389--1407~{\AA} spectral window binned spatially \revise{and spectrally} by two, yielding a resolution of 0.33{\arcsec}\revise{\,pixel$^{-1}$ and 25\,m{\AA}\,pixel$^{-1}$, respectively,} for the Si \textsc{iv} 1402.77\,\AA\,line. 

The high spatial and temporal cadences of this sit-and-stare IRIS observation is well suited to study chromospheric condensation dynamics, since it captures relatively rapid evolution at relatively small scales. In the following, the shape of Si \textsc{iv} line profiles are analyzed using a robust fitting routine, allowing for plasma downflows to be measured. By scanning over the entire ribbon, we were able to find a single pixel [367.2\arcsec, -318.6\arcsec] along the IRIS slit, shown in green in Figure \ref{fig:irisaia}b, containing a condensation event to which we focus our modeling efforts later on.

\subsection{Spectral Fitting} % =========================================
\label{sec:spec}

As discussed in the introduction, previous investigations using Si \textsc{iv} 1402.77\,\AA\ as a diagnostic for condensation have shown asymmetric line profiles to be well modeled by a pair of Gaussians. Likewise, spectra without asymmetries have simply been described by a single Gaussian \citep{li2017}. A preliminary analysis of a variety of line profiles in this work proved a multi-Gaussian approach to be applicable here as well. 

In order for this technique to be viable, however, the spectra need to be produced under optically thin conditions. Si \textsc{iv} 1402.77\,\AA\ is a solar transition region line and has a formation temperature of  $\sim8.5\times 10^4$\,K. Given opacity effects are generally negligible in this region, Si~\textsc{iv} has widely been assumed to be an optically thin line. A simple check of this assumption in our flare ribbon is the comparison of the Si \textsc{iv} doublet lines, where a ratio between $\lambda$1393.75 to $\lambda$1402.77 should be 2:1 ---  the ratio of their oscillator strengths \citep{rybicki1985}. For the condensation event studied in this work, we found a line ratio of 1.96, suggesting the optically thin assumption is accurate in this case.

Plasma properties were determined by fitting the Si \textsc{iv} line at 1402.77\AA.  The line in a given pixel is initially fit with both one and two Gaussians, and error estimates supplied with the level-2 data are used to calculate a reduced chi-squared value, $\chi_\nu^2$, for each of the fits. If $\chi_\nu^2$, falls below a set threshold of 2 for the single-Gaussian fit, that fit is used instead of the double-Gaussian. Additionally, if the peak amplitude of either component of the double-Gaussian is deemed negligible --- below 100\,DN\,s$^{-1}$ --- the fit is rejected in favor of the single Gaussian. Lines whose \revise{wavelength-integrated intensity} fell below 75\,DN\,s$^{-1}$ were not fit at all. 

The double-Gaussian fitting minimizes $\chi_\nu^2$ by varying all six parameters of the two Gaussians, including the center wavelength of each. By allowing the two velocities to move freely, we seek to reduce cross-talk between the two Gaussians, thereby better characterizing two independent plasma components. An initial investigation shows the slower component tending to remain relatively constant over the condensation event, albeit red-shifted to about 10--20\,km\,s$^{-1}$. \revise{We note these persisting downflows agree with previous observations of downflows seen in transition region lines \citep{hansteen1993,doschek1976}, but we do not speculate on its origin here.}
In spite of this red-shift we hereafter refer to the slower component in a two-Gaussian fit as the \textsl{stationary} component. The velocities were bounded between [-60, 500]\,km\,s$^{-1}$ of rest wavelength so as to constrain plasma flows within reasonable values. \revise{In this case, the rest wavelength was calibrated using the IRIS FUV O} \textsc{i} \revise{1355.60~{\AA} line over a quiescent region prior to the flare, with the suggested error estimation being $\sim$1\,km\,s$^{-1}$.} The level-2 data appears to have virtually no background \revise{ --- likely due to the short 4\,s exposure time ---} so none is included in the double-Gaussian fitting. 

The fitting described above was performed across the portion of the eastern ribbon crossing the slit (see Fig.\  \ref{fig:irisaia}). Si \textsc{iv} spectra from IRIS slit pixels corresponding to x= [360.66\arcsec, 376.97\arcsec], were analyzed from time t = 16:36:50 to 17:21:32. Prior to the fitting, the spectrographic data was calibrated with \revise{the \texttt{iris\_prep\_despike} SolarSoft} despiking procedure \revise{to remove bad pixels \citep{freeland1998}}. After the routine was run, we found the largest number (49\%) of ribbon spectra were best fit by a single-Gaussian, while only 12\% were deemed best fit by a double-Gaussian. The remaining 39\% were found to be too faint and were not fit. The centroid of each Gaussian was then used to calculate the Doppler velocity at each pixel-time instance (hereafter instance), with the reference line center for Si \textsc{iv} taken to be its theoretical value of 1402.77\,\AA. Because our routine produced mostly single-Gaussian profiles, we forced continuity between red component velocity measurements at each instance by calling the velocity inferred form single-Gaussian the red component where no double-Gaussian fit was produced.

\begin{figure*}
\includegraphics[width=\textwidth]{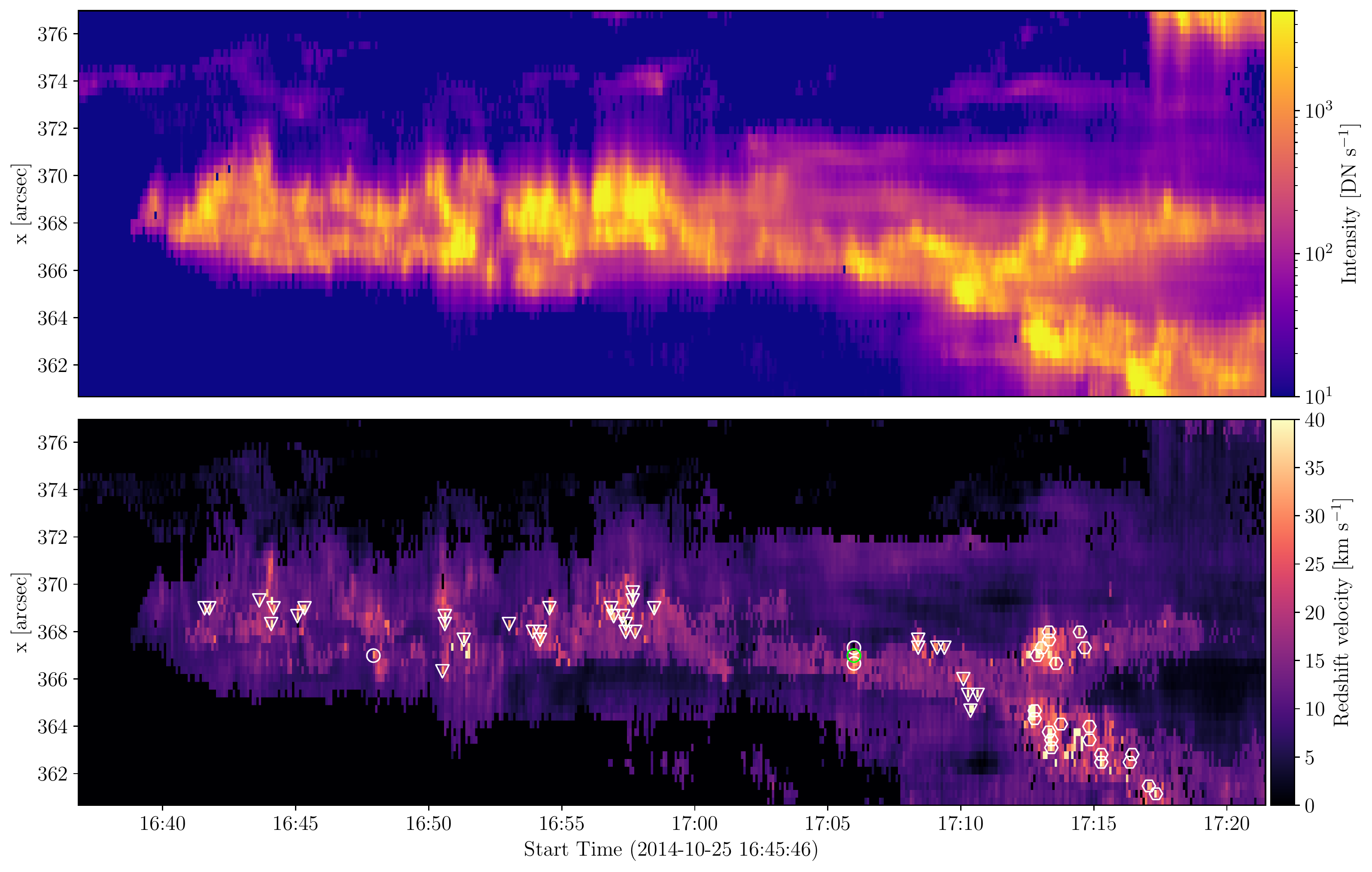}
\caption{Time-distance stack plot showing temporal evolution along a portion of the slit in intensity (top) and red-component velocity (bottom) for Si \textsc{iv} 1402.77\,\AA. The identified condensation events are enclosed according to shape. Circles correspond to archetypal-type condensation events, hexagons and triangles correspond to persisting- and small-type events, respectively. The green circle is the selected instance.
\label{fig:stacks}}
\end{figure*}

The distribution of line properties across space and time is summarized in Figure \ref{fig:stacks}. The top panel shows the intensity of the Si \textsc{iv} line in DN\,s$^{-1}$ as measured by the maximum value in the line profile at each instance. The bottom panel shows velocity measurements inferred from the red components of the double-Gaussian fits, where positive velocity indicates downward flowing plasma along the line-of-sight (LOS). The general evolution of the Si \textsc{iv} spectra shows a ribbon that remains mostly stationary, bound between $\sim$[366\arcsec, 370\arcsec], until it begins to sweep westward (downward) at approximately 17:05 and appears to bifurcate around 17:12.
The structure of the red component velocity appears to follow the evolution of intensity over the evolution of the ribbon, where regions of increased redshift correspond to brightenings in Si \textsc{iv}. Notably, distinct velocity spikes reaching upwards of several tens of km\,s$^{-1}$ stand out among the general downward plasma flows seen, which could be indicative of condensation events. 

To determine which of the spikes seen in Figure \ref{fig:stacks} represents condensation, an identification process was developed according to the principle criteria for condensation: peak downflow velocities on the order of tens of km\,s$^{-1}$ that quickly decay back to pre-peak levels within 60\,s. Using the fitting results, a list of instances having red component velocities above 30\,km\,s$^{-1}$ was compiled. This threshold peak velocity,  more than 3$\sigma$ above the average red component velocity of $9.4\pm6.0$\,km\,s$^{-1}$,  was chosen to clearly differentiate between possible events and more gradual variation of the persistent downflow.  Each of the 126 instances meeting this criteria was visually inspected to verify it conformed to expected condensation evolution. Instances where the peak velocity was a single-frame spike above a constant redshift were rejected.   Instances showing a sustained redshift with no discernible decay profile were likewise rejected. We note that although these instances were rejected, we cannot definitively say the redshifts were not condensations, but only that they did not meet our definition of expected condensation behavior. It is possible condensations might exhibit this atypical behavior, but determining so is beyond the scope of this work. 

Through the identification process we found 60 condensation events over the lifetime of the flare ribbon. Further inspection of the spectral lines in each event motivated us to classify the condensations into three types: persisting, small, and archetypal. Each condensation is identified by shape according to event type in Figure \ref{fig:stacks}. Persisting events were shown to exist among a larger body of velocity spikes, where the redshifts appear to have an envelope that is itself decaying over the course of several minutes. Small events exhibited red component intensities much smaller than their stationary counterparts. At peak downward velocity, the average ratio between the two components in these cases was 8.8\% and grew as the condensation progressed.  The nature of these two types of events are discussed in more detail within an appendix.
 
\revise{Archetypal events were those that best matched the prevailing theoretical understanding of chromospheric condensation: downflows with a peak flow velocity that decays on the order of seconds \citep[i.e.][\ALpaper]{fisher1989}. Furthermore, from a phenomenological perspective, many of the lines used to observe condensation behavior have a distinct red-asymmetry in the profile from which the velocity is measured (and defined by). Together, these properties constitute the ideal, archetypal definition of chromospheric condensation.} These events are in direct contrast to the other two types because they have clear spectral redshifts whose evolution did not belong to \revise{an overarching velocity envelope}, and thus have more exemplary condensation behavior. \revise{However, although our definition of this type might imply they would be the more prevalent type throughout the flare ribbon,} only four such events were found. \revise{Three of these occurred} concurrently, spanning consecutive pixels. \revise{Even though the small number of archetypal events found is perhaps unexpected, we do not speculate on the possible reasons in this work.} For completeness, we repeated the entire fitting process using instances whose redshifts were derived from single-Gaussian profiles, rather than the red component of double-Gaussian profiles. However, no such events were found.

\begin{figure*}
\includegraphics[width=\textwidth]{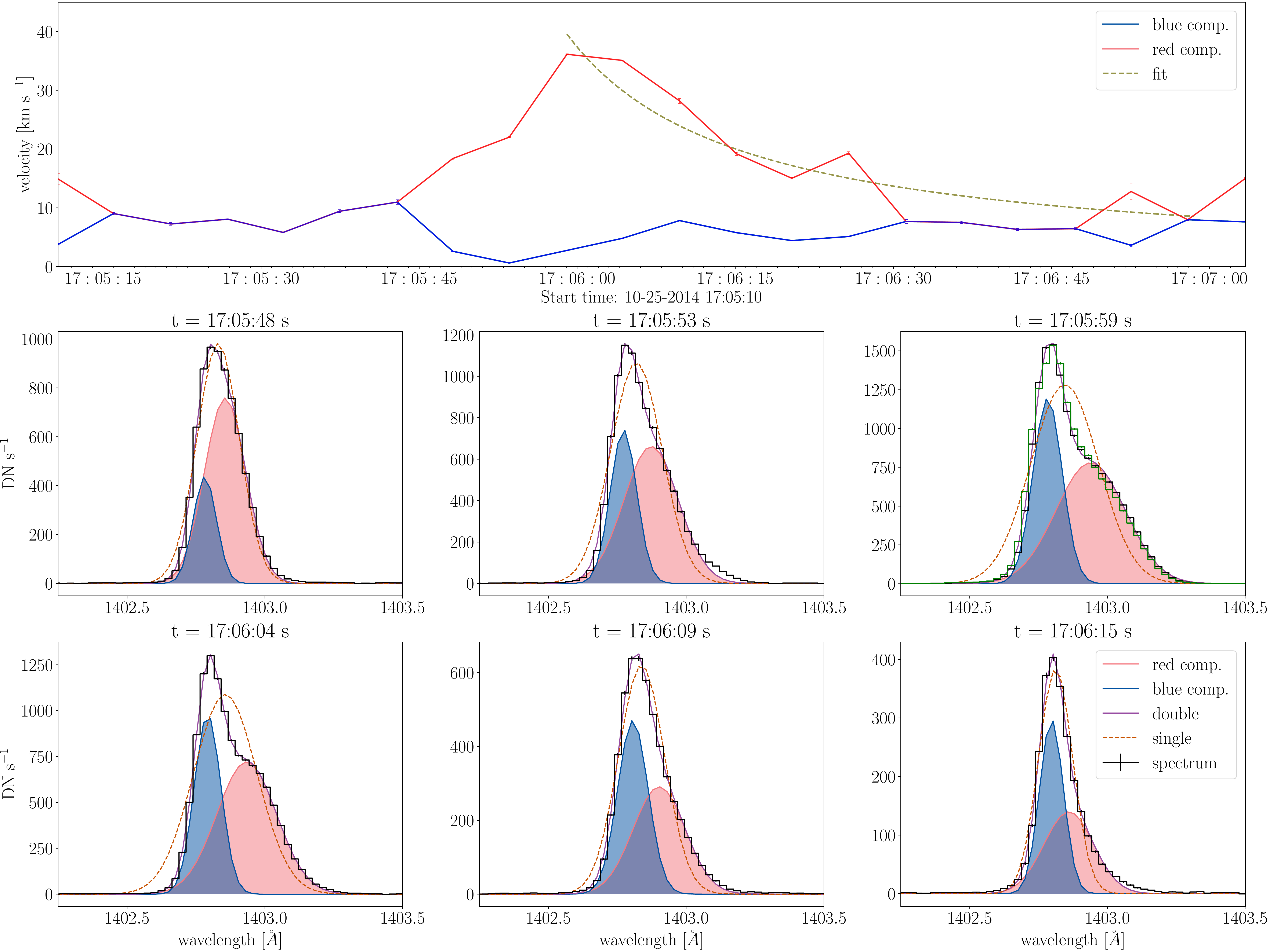}
\caption{Evolution of the condensation event. Top panel: Velocity time series of the condensation. Red and blue points indicate the red and stationary components resulting from the double-Gaussian fit to the spectral data, respectively. Times with a single point indicate instances where a single-Gaussian fit was sufficient. The best fit of Equation (\ref{eqn:ut}) to the red component is given by the green dashed curve. Bottom panel: Selected IRIS Si \textsc{iv} 1402.77\,\AA \ spectra (black curve) are shown along with their red and stationary components from the fit to a double-Gaussian (red and blue solid curves, respectively). The combined profile from the double-Gaussian fit is overlaid in purple. The rejected single-Gaussian fit (orange dashed) is shown for comparison. \revise{The green curve at t=17:05:59 shows the corresponding Si \textsc{iv} 1393.76~{\AA} line profile divided by 1.96, the line ratio between the two spectra}.
\label{fig:px69}}
\end{figure*}

Of the four archetypal events identified, the one with the clearest condensation behavior is shown in Figure \ref{fig:px69}. The velocity time series is shown in the top panel. Prior to the condensation, the Si \textsc{iv} has a single-Gaussian profile and is continuously redshifted at roughly 10\,km\,s$^{-1}$. Starting at $\sim$17:05:45, the red component quickly increases from this constant downflow and reaches its peak velocity of 36\,km\,s$^{-1}$ at 17:05:59. The condensation then evolves characteristically, decaying from peak velocity down to the pre-condensation, stationary redshift in just over 30\,s. During this time, the stationary component largely matches the sustained redshift we see before and after the event. The instance at peak velocity is indicated by a green circle on Figure \ref{fig:stacks} for reference. 

What makes this condensation an archetypal event is not the velocity evolution, however, but the degree of redshift in the line profiles. The bottom panels of Figure \ref{fig:px69} illustrate the evolution of the spectral line around the time of peak velocity. Here we see the red component -- already enhanced to nearly 800\,DN\,s$^{-1}$ -- shift rightward until a clear red-asymmetry is formed. This asymmetry is in part due to the stationary component, which remains comparable in strength to the red counterpart throughout the event. After reaching peak velocity, the two components fade together as the redshift quickly recedes back to the stationary velocity. The red component width was also found to be larger than the stationary component, with a maximum width of 27\,km\,s$^{-1}$ occurring at peak velocity. 

Not only is this event the best example of condensation found in the flare, but there are also other aspects that make it distinctive. First, it exists in the center of the three clustered events pointed out above. We found no other condensations having this quality. 
Second, there exists \revise{three instances of} condensation on the same pixel as our event, with our event being the second. The peak of the earlier event occurs at 16:48:06, nearly 20 minutes prior. Given the time scale of energy deposition driving the \revise{first} condensation is thought to be much less than 20 minutes and the \revise{IRIS} pixel is assumed to be the footpoint of a stationary flare loop, the earlier \revise{instance} suggests the \revise{loop may have been impulsively heated once before our event}. The implications of a flare loop undergoing multiple energizations is discussed in Section \ref{sec:dis}. Notwithstanding these aspects, we found the event to be an exceptional example of condensation behavior. For this reason, we focus our study on this event for the remainder of this work.

\subsection{Dynamics} % ===============================================
\label{sec:dynamics}

\ALpaper\ used numerical experiments to establish a relationship between the pre-reconnection density scale height of the chromosphere, $H$, and the properties of condensation. These experiments were designed to be relatively simple, with steady energy transport by thermal conduction into an isothermal, stratified chromosphere. An analytical model was developed from the simulation results, which predicted that the speed of the condensation propagating into the chromosphere would be a power-law dependent on the upstream chromospheric density: $v_s \sim \rho^{-\alpha}$. The numerical solutions, in conjunction with the analytical model, found the decay of the condensation to be well described by both the peak downward fluid velocity $u_0$ and the velocity-half life $\tau$. An expression for the evolution of the condensation fluid velocity is given by 
\begin{equation} % u(t)
    \label{eqn:ut}
    u(t) = \frac{u_0}{1+ t/\tau}.
\end{equation}

In this work, we apply the findings of the initial, theoretical investigation to our condensation observation. Starting at peak downward velocity at 17:05:59, we fit the red component velocity with Equation (\ref{eqn:ut}), with $u_0$ and $\tau$ as free parameters. The best fit is plotted as a dashed line in the top panel of Figure \ref{fig:px69}. Here the fit infers a peak peak velocity $u_0$=39\,km\,s$^{-1}$, slightly higher than its measured value, and a half-life of $\tau$=16\,s. Given the degree to which the velocity evolution follows the fitted curve, we conclude the decay of the condensation event is well approximated by the expression predicted from the previous study. 

One of the conclusions of \ALpaper\ was that the parameters of Eq.\ (\ref{eqn:ut}) were related to the stratification in the pre-flare chromosphere. A parameter investigation led to an empirical relationship between scale height and the two condensation properties, where the product of the peak velocity and redshift half-life was found to be proportional to the scale height as $H \approx 0.6 u_0\tau$. Using the values of $u_0$ and $\tau$ from our fitting, we determined the condensation must propagate into a chromosphere initially stratified with density scale height $H = 369$\,km.  In light of the fact that the pixel is surrounded by a flare ribbon, and that the same pixel experienced an earlier instance of condensation, it is probably not reasonable to expect the pre-energization chromosphere to be in pristine equilibrium.  In light of this, the value $H=369$\,km seems plausible. 

A second conclusion of \ALpaper\ was that when driven by modest conductive flux, $F$, \revise{(i.e. $F \lesssim 2 \times 10^{10}\,{\rm erg\,cm^{-2}\,s^{-1}}$)} the peak downflow speed, $u_0$, follows the scaling with condensation velocity proposed by \citet{longcope2014}
\begin{equation} % u(t)
    \label{eqn:vc}
    v_c = C_c \left(\frac{F}{\rho_{{\rm ch},0}}\right)^{1/2},
\end{equation}
where $\rho_{{\rm ch},0}$ is the initial mass density of the chromosphere and $C_c\simeq10^{-4}\,({\rm cm\,s^{-1}})^{-1/2}$ is an empirical constant.  Equating this expression for $v_c$ with $u_0$ leads to a flux
\begin{eqnarray} % u(t)
    \label{eqn:F}
    F &=& \rho_{{\rm ch},0}\,\left(\frac{u_0}{C_c}\right)^2 = \frac{m_p p_0}{2k_{\rm b}T_0}\,
    \left(\frac{u_0}{C_c}\right)^2 \nonumber \\[7pt]
    &=& p_0 \times (10^9\, {\rm cm/s}),
\end{eqnarray}
where $p_0$ is the pre-event chromospheric pressure, and $T_0=10^4$\,K is the temperature of the pre-event chromosphere.  To obtain the final expression we used the 
 the observed downflow velocity, $u_0$=39\,km\,s$^{-1}$.
 
 This second relation provides a constraint, but not a precise value for the energy flux.  We might consider a range of plausible pressure for a loop embedded in a flaring active region, but not yet energized itself.  Taking that range to be 1--10 ${\rm erg\,cm^{-3}}$ yields fluxes in the range 
 $10^{9}$--$10^{10}\,{\rm erg\,cm^{-2}\,s^{-1}}$, which are also plausible, and modest enough to justify our initial assumption.  We used different observational methods, described in the next section, to obtain a value of $F$ which turns out to fall within this range.

\section{Energy Deposition} \label{sec:energy}

The condensation event described above occurs at one footpoint of a loop energized by a flare.  In order to model the condensation, we seek to model the entire energy release process into that loop.  Here we derive the energy input for the model we use for the entire flare loop.

\subsection{Energy Transport} \label{sec:hxr}

The relationship between chromospheric response and coronal energy release will depend on the mechanism transporting that energy \revise{\citep{fletcher2011,holman2011,kontar2011}}.  According to one well-known theory, energy is transported from the corona \emph{via} non-thermal electrons.  While the bulk precipitation of these electrons directly heats the chromosphere by Coulomb collisions, a very small fraction scatter and produce bremsstrahlung radiation typically seen as HXR emission. Observations of HXRs that spatially coincide with flare ribbons are taken to be evidence of this transportation process \citep{brown1971}.

\begin{figure}[t!]
\includegraphics[width=\columnwidth]{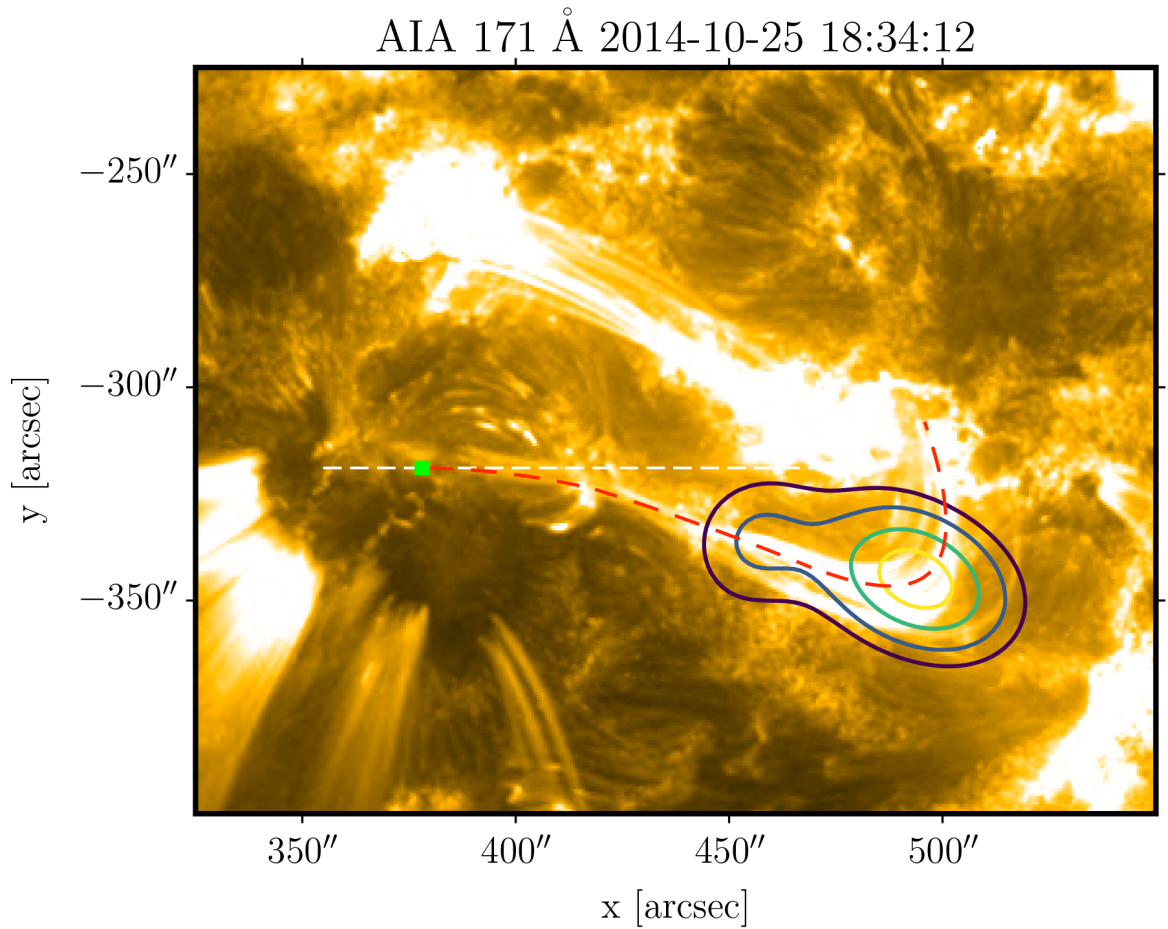}
\caption{AIA 171\,\AA\,image with RHESSI 25--50\,keV contours overlaid at [0.35, 0.50, 0.70, 0.90] of maximum emission, respectively. The contours correspond to emission integrated over 16\,s from 17:05:45 to 17:06:01 and are shifted to coalign with the AIA image. The IRIS slit (white dashed line) and the condensation pixel (green marker) are also shifted from the time of peak condensation velocity at 17:05:59. The red dashed line illustrates the approximated location of the flare loop with a footpoint matching the condensation pixel. 
\label{fig:aiarhessi}}
\end{figure}

HXR data for this flare was collected using the Reuven Ramaty High Energy Solar Spectroscopic Imager \citep[RHESSI;][]{lin2002}. Figure \ref{fig:aiarhessi} shows the contours of the RHESSI image with energy range 25-50\,keV integrated for 16\,s over the time of peak condensation velocity from 17:05:45 to 17:06:01. Underlying the contours is a AIA\,171\,\AA\, image at 18:34:12 showing several post-flare loops. Analysis of the image identified a loop having a footpoint matching the selected IRIS pixel, shown here as a red dashed line. The contours are shifted so that they coalign with the later AIA\,171\,\AA\, image.  The location of the HXRs along the top of the flare loop structure suggest the emission in 25--50\,keV is confined to the corona, originating from a possible thick-target coronal source \citep{veronig2004,sui2004}. The absence of any flare ribbon emission from the location of the HXR contours, along with the 35\% and 50\% contours appearing to extend along one leg of the loop, add further evidence of a coronal source. Furthermore, spectral analysis of the HXR flux shows a soft non-thermal power-law index of $\gamma \sim8$. This is also indicative of a thick-target coronal source, as non-thermal emission from these sources are associated with a HXR spectra with $\gamma > 6$ \citep{veronig2004,krucker2008,guo2012_1}.

These findings are further supported by previous investigations of this flare. \citet{kleint2017} measured an index of $\gamma \sim7$, and also proposed the possibility of a thick-target coronal source for the HXR emission. Their work, however, points out the effects of pile-up that occurred during the RHESSI observation. As the artifacts produced from pile-up tend to harden non-thermal HXR spectra, we cannot speculate on the details of the coronal HXR emission in this work. In contrast, \citet{kowalski2019} measured a similar power-law index of $\gamma = 8$--$9$ using Fermi/GBM data, which did not suffer from pile-up. 

The location of non-thermal emission at the apex of the identified loop, and not in the vicinity of the IRIS footpoint pixel, strongly suggests the condensation event we have identified is not a response to energy transported to the chromosphere via electron precipitation.  Instead we propose that energy is somehow thermalized somewhere in the corona, and transported to the chromosphere by thermal conduction.  This scenario differs from methods often used in 1D models of flares --- that is, chromospheric heating delivered via electron beam as seen in flare codes such as RADYN \citep{carlsson1997,allred2015} and HYDRAD \citep{bradshaw2003,bradshaw2013}.  For our model we estimate the energy deposited into the identified flare loop by converting UV emission measured at the loop footpoint into a heat flux. The details of this method are outlined below. 

\subsection{Computing Energy Input {\em via} UFC Modeling} \label{sec:UFC}

Having established the energy flux to be conductive, we quantify
the time-dependent energy flux itself using the UFC method.  First developed in \cite{qiu2012} and \cite{liu2013}, the method assumes that the energy flux (in units of erg\,cm$^{-2}$\,s$^{-1}$) deposited at each ribbon pixel is proportional to the brightness of the UV 1600~{\AA} light curve (in units of DN~s$^{-1}$ pixel$^{-1}$) there. A scaling constant between AIA 1600~{\AA} lightcurve at each flaring pixel and its related heating flux is obtained by comparing coronal observations in a set of SXR and EUV bands with synthetic emissions produced from an efficient zero-dimensional loop evolution model EBTEL \citep{klimchuk2008}. The empirical relation for energy input is adjusted to achieve the best agreement across all bandpasses.

\revise{To determine how emission in AIA 1600~{\AA} relates to the emission of our condensation event, we first coalined the two data sets. The process was accomplished using SolarSoft mapping software, where a pair of single image maps of AIA 1600~{\AA} and IRIS SJI 1330~{\AA} were generated and used as input for a two-dimensional cross-correlation analysis provided by the \texttt{get\_correl\_offsets} routine. We found the SJI image at 17:06:12 to be the best base image, as it was temporally close to the peak condensation time and had the closest time to an AIA image at 17:06:16. (We point out that these are the same images shown in Figure \ref{fig:irisaia}). Because AIA has a different spatial resolution and FOV than SJI, the 1600~{\AA} image was rebinned and cropped to match the 1330~{\AA} image. The cross-correlation between the two images produced a spatial offset of x,y=[-0.52\arcsec, -2.02\arcsec] by which the entire AIA dataset was shifted to correctly align with IRIS.}

\begin{figure}
\centering
\includegraphics[width=0.5\textwidth]{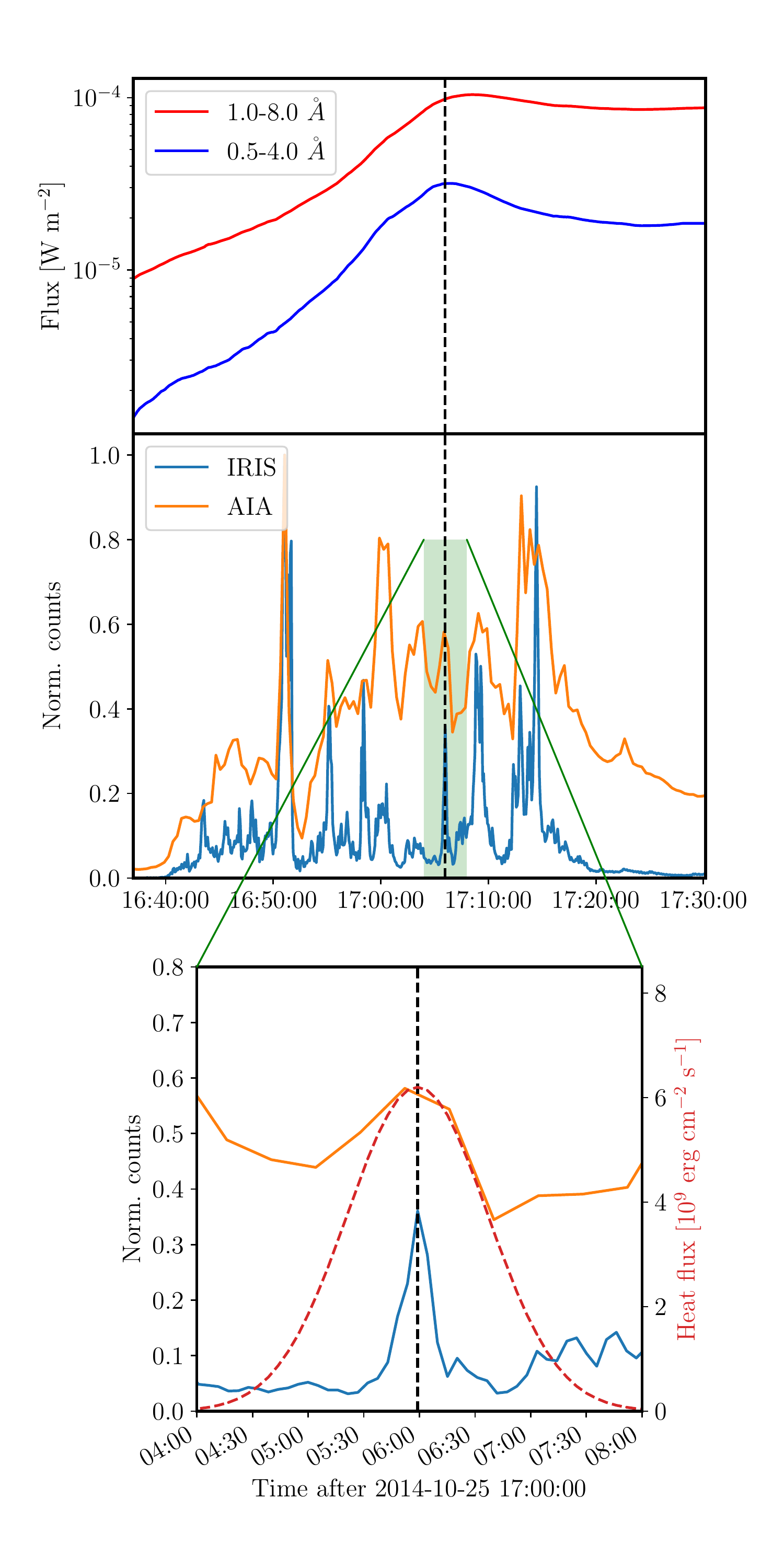}
\caption{Light curves of the 2014 October 24th X1.0 class flare. GOES X-ray bands are shown in the top panel. The middle panel shows coalined AIA 1600~{\AA} and IRIS Si \textsc{iv} 1400--1405~{\AA} light curves for the selected pixel. The period corresponding to our condensation event is enlarged in the bottom panel.  The red dashed line is the heat flux derived from the UFC method, read against the axis on the right. The black dashed line throughout is the moment of relative peak emission in Si \textsc{iv} signifying our event.
\label{fig:pxlcs}}
\end{figure}

The middle panel of Figure \ref{fig:pxlcs} shows the normalized\revise{, coaligned} light curves for our pixel over the lifetime of the flare in both AIA\,1600\,\AA\, and IRIS Si \textsc{iv} integrated over the range 1400--1405\,\AA. The black line denotes the time of peak downflow of the event. The two GOES soft X-ray channels in the top panel, show that our condensation event occurs at the very end of the impulsive phase of the flare.  An enlargement of the intensity peaks corresponding to our condensation event is shown in the bottom panel. The concurrent brightening of both light curves demonstrates how the heating mechanism driving the condensation seen in Si \textsc{iv} is also reflected in the 1600\,\AA\, channel. Therefore, by implementing UFC on the singular brightening in 1600\,\AA\,, we can infer the energy flux driving our condensation.

\begin{figure}[tp]
\centering
\includegraphics[width=0.5\textwidth]{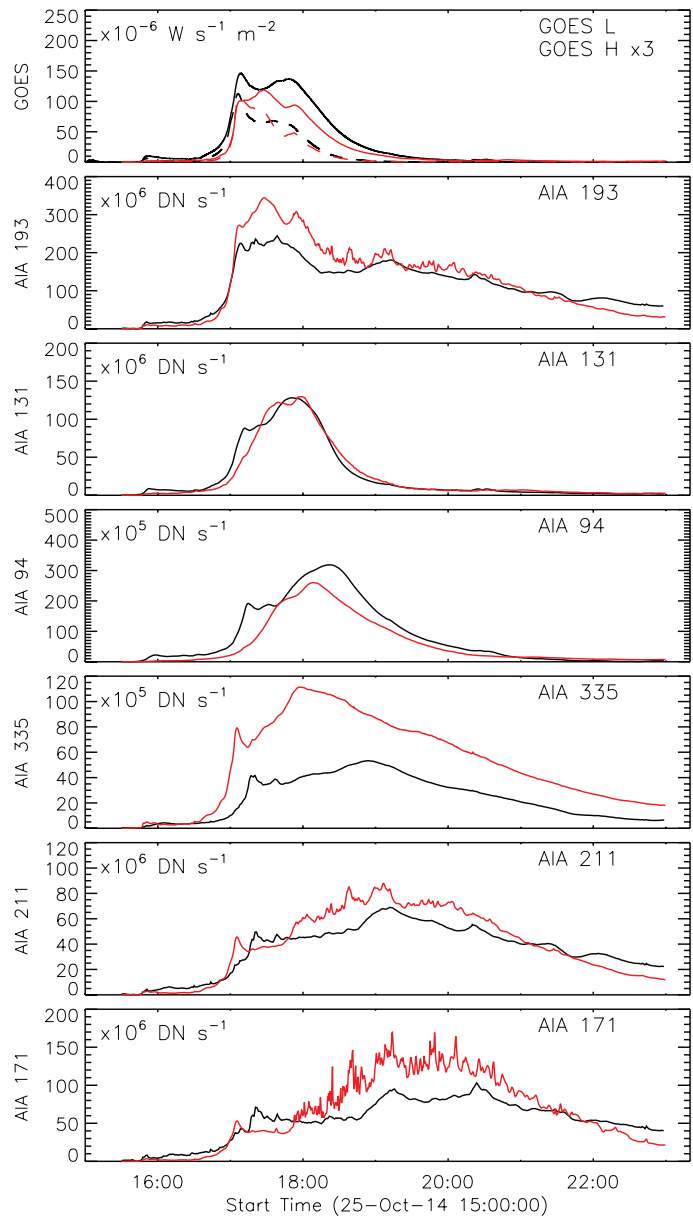}
\caption{Coronal response for the entire flare seen in six AIA channels and the two GOES bands (black). The best match to the observed light curves produced from the 0D EBTEL model are shown in red.
\label{fig:ebtllcs}}
\end{figure}

The detailed setup of the UFC method can be found in previous studies on this method (\citealp{qiu2012, liu2013, zhu2018}).
Figure \ref{fig:ebtllcs} displays the best match between the total synthetic (red) emission and the observed count rate (black) for each respective channel, giving an average value of the correlation coefficients of 0.96. Although the agreement between the synthetic and  measured  coronal  flare  response is adequate, we note the synthetic AIA 335~{\AA} emission is $\sim$2.5 times larger than the observed values. This discrepancy has been seen in previous studies, albeit with the model lightcurves having insufficient emission levels possibly due to low temperature plasmas of less than 1\,MK not reproduced by the EBTEL model \citep{qiu2012,qiu2016}.

The scaling constant (denoted by $\lambda$) determined from the full-flare comparison is used to compute the energy flux delivered to the single-pixel condensation event. The single 1600~{\AA} peak is fit with a Gaussian function, which is then scaled by $\lambda$ to produce heat flux $F(t)$. The agreement in Fig.~\ref{fig:ebtllcs} suggests a value of  $\lambda$ at $5 \times 10^6$~ erg~cm$^{-2}$~DN$^{-1}$ for this X1.0 flare. The resulting heat flux has a Gaussian profile ($\sigma$ = 38\,s) as denoted by the red-dashed curve in the bottom panel of Figure \ref{fig:pxlcs}, read against the right (red) axis. The flux deposits a total heat energy per unit area of  $\int F dt = 6 \times 10^{11}$\,erg\,cm$^{-2}$ and reaches a peak of $F_p=6.2 \times 10^9$\,erg\,cm$^{-2}$\,s$^{-1}$. 

\section{One-dimensional Flare Loop Modeling} \label{sec:1D}

The energy input inferred above is used in a one-dimensional flare loop model to compare to the observed chromospheric condensation.  The properties of the loop are found from 
a constant-$\alpha$ field extrapolation from the line-of-sight magenetogram from Helioseismic and Magnetic Imager \cite[HMI;][]{schou2012}, shown in Fig.\ \ref{fig:irisaia}(c).   A value of $\alpha = - 0.0055$\,Mm$^{-1}$ produces a 85\,Mm long field line, shown as a violet curve, with its negative footpoint roughly at the the pixel location.  The loop extends beyond the nearby positive polarity to a footpoint just north-west of the slit's end.  This is similar to the footpoint we observed in the 171 \AA\ loop shown in Figure \ref{fig:aiarhessi}, so we believe the our computed loop length to be reasonable.   The force-free field line does not, however, curve as far south as the 171 \AA\ loop. Although this discrepancy is likely due to the actual field not having a uniform $\alpha$, the model is taken to be an acceptable approximation of the loop.

The chromospheric response to the UFC-derived energy release is modeled using the same one-dimensional, hydrodynamic simulation used in \ALpaper. Following those numerical experiments, a rigid magnetic flux tube is initialized in mechanical equilibrium with full length $L$=85\,Mm determined from the aforementioned field extrapolation. Appended to the ends of the loop is a simple, isothermal chromosphere of temperature $T_{ch,0}$=10,000\,K. The pre-flare chromosphere is gravitationally stratified according to the density scale height $H$, such that the density increases exponentially with depth into the solar atmosphere. We set the gravitational acceleration $g$ in order to make the scale height in our model match $H=369$\,km, as was inferred in Section \ref{sec:dynamics} from IRIS observations. The coronal region of our flux tube is initialized in equilibrium by solving the RTV relations set by our apex loop temperature,  T$_{co,0}$ \citep{rosner1978}. To ensure continuity between the corona and chromosphere, the pressure and and temperature are continuous across the boundary between the regions ($z=0$). We set T$_{co,0}=2$\,MK, which produces a pre-event chromospheric pressure $p_0=0.7$\,erg\,cm$^{-3}$ at the boundary.

The initialized flare loop was then evolved in time using the \textsc{PREFT} numerical code \citep{longcope2015}. Run in the straight tube configuration, the code solves one-dimensional gas dynamic equations on a Lagrangian grid, where cells have constant mass per unit flux. At $t$=0\,s, the heat flux inferred in the previous section is deposited along the top of the loop in a spatial tent profile centered at the apex and spanning half the loop length. In order for the Gaussian heating function to have a finite duration, we set the peak of the heating to occur at $t=150$\,s. The simulated flare energy release quickly forms a conduction front in the corona, which transports energy towards the loop footpoints. Due to the gradual start of the Gaussian-profile heating, the front first reaches the chromosphere at $t=72$\,s into the energy release, 78\,s prior to peak energy flux.

\begin{figure}[t!]
\centering
\includegraphics[width=0.7\textwidth]{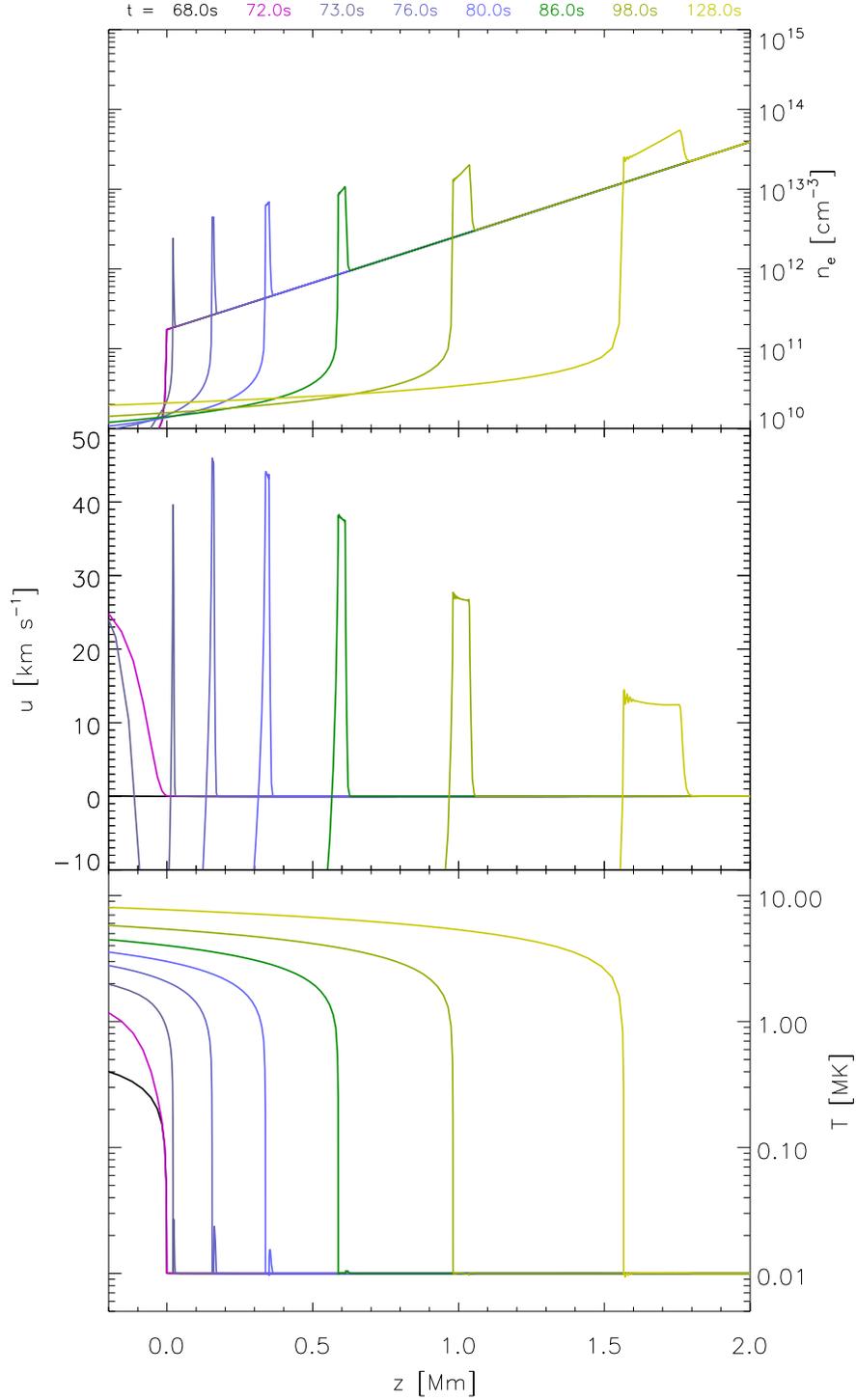}
\caption{Chromosphere evolution from the \textsc{PREFT} model results. Panels show density, fluid velocity, and temperature, from top to bottom, respectively. Positive velocities represent downflows. Each panel is plotted against $z$, with $z$ = 0 being the top of the chromosphere and increases with depth. Colors correspond to different simulation times, with t=68.0\,s showing the pre-condensation chromosphere.
\label{fig:preftraw}}
\end{figure}

The energy deposition into the chromosphere then drives condensation in the form of a downward propagating hypersonic shock. Snapshots of the plasma evolution in the chromosphere at different times during the condensation are shown in Figure \ref{fig:preftraw}. The initial chromospheric conditions can be seen at $t$=68\,s. At $t$=72\,s the conduction front has reached the top of the chromosphere, defined here by the coordinate $z$ such that $z$ = 0 corresponds to the boundary between the chromosphere and the corona. Over the next several seconds, the shock front rapidly forms as the downward fluid velocity --- defined to be positive in this case --- increases to flows of up to $\sim$45\,km\,s$^{-1}$. After reaching peak velocity at $t=76$\,s, the condensation begins to decay as it encounters ever increasing density in the chromosphere. By $t$=128\,s, the strength of the condensation has decayed considerably to speeds of ${\sim}10$\,km\,s$^{-1}$. \revise{We note the ringing pattern seen in the velocity panel at $t=128$\,s is a numerical artifact, likely arising from numerical dispersion of waves at  grid scales.  These are damped quickly and do not compromise conserved quantities.}

In contrast to the simulations run in \ALpaper, we did not impose a limit on the electron number density in calculating radiative losses from the plasma. The limit was originally designed to eliminate artificial radiative instabilities that arose when high values of flare energy flux were left on for the duration of the simulations. We found the energy flux delivered to the loop in this work to be small enough that it did not result in these instabilities, and thus omitted it here. In doing so, the radiative losses became more efficient downstream of the condensation front and the shock developed into what is referred to as a \textsl{radiative} shock \cite{fisher1986_2}. This effect can be seen during the early states of the condensation in Figure \ref{fig:preftraw}, where the width of the stock itself is only several km thick. Furthermore, with respect to the analytical model derived in the previous work, we found the power-law relationship between shock velocity and upstream chromospheric density to also be true in the case of a radiative shock. Thus the theoretical findings in that work, namely the predicted evolution of condensation discussed in Section \ref{sec:dynamics}, can still be applied here. 

The simulation also \revise{supports the observation} that the Si {\sc iv} line is optically thin. Optical depth of the line is bounded from above by integrating the number density $n$ of the Si \textsc{iv} ion over the entire line of sight
\begin{equation}
  \tau_{1403} ~=~ \sigma^{(1403)}_0 \, \int\, n\, d\ell ~~,
\end{equation}
where $\sigma^{(1403)}_0 = (3.78\times10^{-8}\,{\rm cm^3\,s^{-1}})/\, \Delta v$ is the absorption cross section at the center of a line broadened to velocity $\Delta v$.  Assuming Si at photospheric abundance and equilibrium ionization, and taking $\Delta v=20$\,km\,s$^{-1}$, an integral over the entire left leg of the simulated loop yields $\tau_{1403}$ increasing during the condensation, but never exceeding $0.04$.  The intensity ratio of the 1393\AA\ line to this one,
\begin{equation}
  {I_{1393}\over I_{1403}} ~=~ 1~+~e^{-2\,\tau_{1403}}~\simeq~2~-~2\,\tau_{1403} ~~,
\end{equation}
would not fall below $1.92$ under the simulated conditions.  This value is consistent with our observation.

\subsection{Synthetic Profiles}

Because the Doppler velocity measured by IRIS in Section \ref{sec:dynamics} only reflects the plasma flow of Si \textsc{iv} in the temperature range where the ion is most abundant, the results of our 1D simulation cannot be directly compared to observation. To make this comparison, we thus created a series of synthetic Si \textsc{iv} spectra using the simulation results.

The synthetic Si \textsc{iv} lines were computed under the assumption of optically thin formation conditions, given the results of our simulation. Under this circumstance, the contribution function used in calculating emergent intensities was constructed using the CHIANTI v.8 atomic database \citep{chianti}. However, CHIANTI computes intensity under the assumption of ionization equilibrium, which may not be justified in the rapidly changing plasma conditions produced by the hypersonic shock \citep{shapiro1977,dudik2017}. In this event, the plasma is considered to be in non-equilibrium ionization state and requires a solution to the time-dependent ionization balance equation in order for the synthetic line intensity to be accurate. The change in fractional ionization, $f_i$, for a single Lagrangian simulation cell over time is found by
\begin{equation}
        \frac{df_i}{dt} = n_{e}\Bigl[C_{i-1}f_{i-1} -(C_i+R_i)f_i + R_{i+1}f_{i+1}\Bigr],
\end{equation}
where  $n_e$ is the electron number density. The ionization and recombination rates are given by $C$ and $R$, respectively, and are also constructed using CHIANTI \citep{shen2013}. The set of ionization fractions for the entire simulation are then normalized to the corresponding equilibrium ionization state to produce a dimensionless factor, $\zeta$, by which we scale the original contribution function. This in effect transforms our original contribution function, such that $\tilde{G}(T) = \zeta G(T)$, accounting for non-equilibrium ionization effects.

For each time in the evolution of the flare loop, the intensity from a single cell was calculated using the emission measure of that cell, $EM_i$, and the transformed contribution function,
\begin{equation}
    I_{i} = \ EM_i \tilde{G}(T_i).
\end{equation}
The $EM_i$ was calculated as the product between \revise{$n_e^2$} and length of the cell. This single-cell intensity is then used to create a synthetic Si \textsc{iv} line together with the normalized line profile, $\phi_i(\lambda)$, for that cell. Here the line profile is a Gaussian centered at the wavelength Doppler shifted by the axial velocity in the cell. The width of the line is determined by a quadrature sum of thermal, $\sigma_{th}$, non-thermal, $\sigma_{nt}$, and instrumental widths, $\sigma_{inst}$. Thermal broadening is calculated by the Doppler width, defined by the temperature of each cell as $\sigma_{th} \sim \sqrt{T}$. Non-thermal and instrumental broadenings are set to be $\sigma_{nt}=20$\,km\,s$^{-1}$  and $\sigma_{inst}=3.9$\,km\,s$^{-1}$, respectively, as \revise{motivated by} \citet{depontieu2015}. Finally, the synthetic lines from each cell are summed along one leg of the loop to create a total emission spectrum at a single instant: $I(\lambda) = \sum_i I_{i} \phi_i(\lambda)$.

\begin{figure*}[t]
\includegraphics[width=\textwidth]{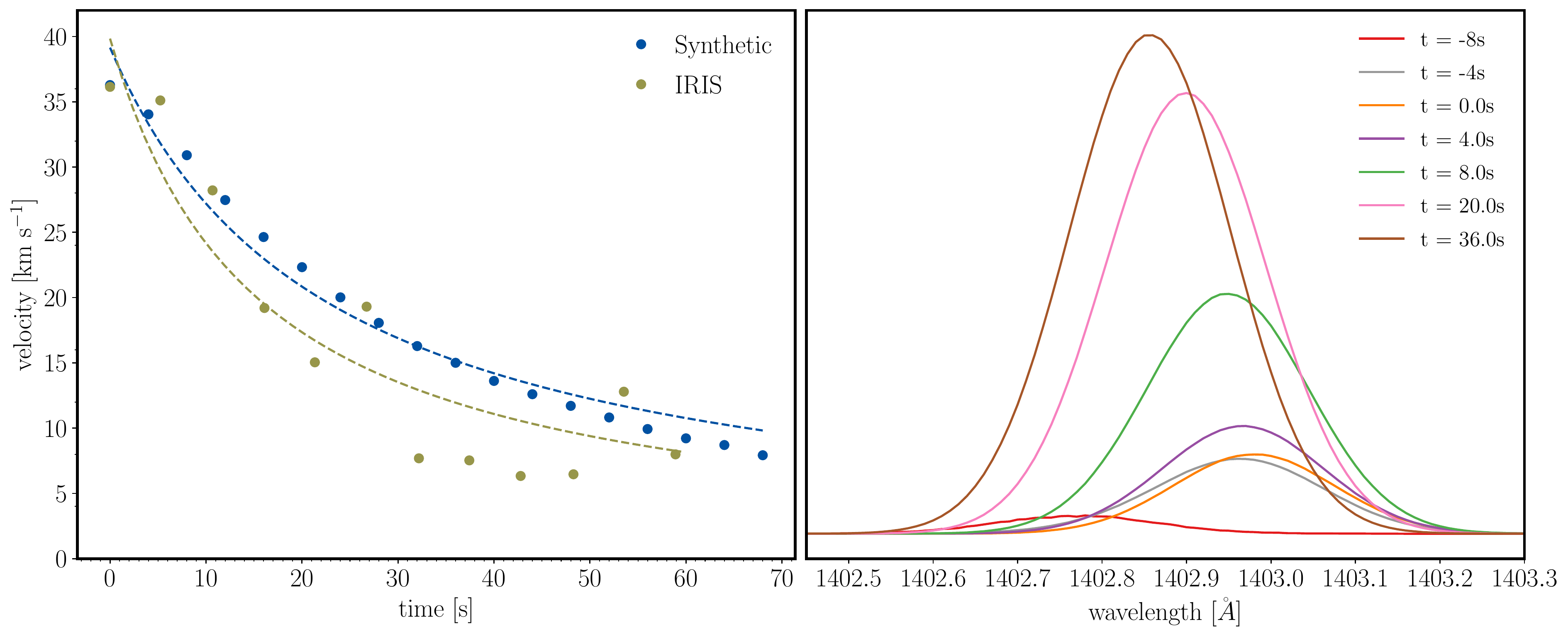}
\caption{Results from the \textsc{PREFT} model. A time series of the velocity measured from the synthetic spectra (blue) is shown in the left panel, along with the red component of the IRIS velocity observation from Figure \ref{fig:px69} (green). The dashed lines show the respective fits to the time series according to Equation \ref{eqn:ut}. The right panel shows the temporal evolution of the time-averaged synthetic Si \textsc{iv} 1402.77\AA\,spectra at several times, with $t$=0\,s corresponding to the time of peak downflow velocity in the simulation. Each spectrum is normalized by the peak amplitude of the pre-condensation line profile, shown as the red $t=-8$\,s line. The pre-condensation profile has been scaled by a factor of 20 for clarity. 
\label{fig:synthpreft}}
\end{figure*}

Synthetic Si \textsc{iv} lines were calculated at each 0.2\,s time step over the duration of the simulation. These were then averaged over the $4$\,s exposure time of the IRIS SG in order to create spectrum with an analogous evolution to those measured from observation. A selection of these spectra are shown in the right panel of Figure \ref{fig:synthpreft}. The given times have been shifted such that $t$=0\,s corresponds to the moment of peak downflow velocity. The initial pre-flare intensity, scaled here by a factor of 20, is shown at $t=-8$\,s. At $t=-4$\,s, the condensation begins to form as the line profile shifts rightward and increases in intensity. Once the condensation reaches peak velocity, it begins to decay as the profiles shift leftward over time. Peak intensities of the spectra increase continuously over the course of the condensation\revise{, reaching values almost an order of magnitude higher than what is seen observationally.  The synthetic intensity is computed assuming a filling factor of unity within the pixel.} 

The intensity increase of our synthetic lines is contrary to the line profiles seen in the IRIS observation, where the evolution in the line profile intensity roughly follows the evolution of the condensation velocity. As seen in the bottom panel of Figure \ref{fig:pxlcs}, the peak in Si \textsc{iv} intensity occurs simultaneously as the peak in condensation velocity at 17:05:59. Moreover, the routine used to synthesize the spectrum failed to create a two-component Gaussian profile, as was seen in observation. Possible reasons for these discrepancies are discussed in the following section. 

Each synthesized spectrum was then fit with a single Gaussian in order to derive a velocity time profile for the condensation produced by the simulation. Given the location of the selected IRIS pixel, roughly $\theta=30^\circ$ from disk center, we scale each synthetic velocity measurement by $\cos\theta=\sqrt{3}/2$ to account for LOS effects. The resulting velocity evolution is shown in the left panel of Figure \ref{fig:synthpreft}. We characterize the evolution by fitting the time series with the theoretical expression for condensation decay given by Equation (\ref{eqn:ut}), with the best fit illustrated by the blue dashed curve. Here the velocity reaches a peak of $36$\,km\,s$^{-1}$ and has a half-life of $\tau=22$\,s. The evolution of the red component velocity inferred from the IRIS observation in Section \ref{sec:dynamics}, starting from the instance of peak velocity, is shown alongside the synthetic values. The best fit to Equation (\ref{eqn:ut}) is also shown. Notably, the observed peak velocity matches that of the one synthesized from our numerical experiment.  Although the synthetic condensation takes slightly longer to decay than what is seen observationally, the model does an adequate job overall at reproducing the velocity evolution of the selected event.

An unexpected result from the simulation was the relationship between the flare heating profile and the lifetime of the modeled condensation. We found the heat flux subjected to the model loop lasting five minutes produces a condensation lasting for roughly a third of that time. After the heat deposition is turned on, the condensation rapidly forms in the chromosphere after a 72\,s delay. At this time, while the flare energy release is still increasing in strength, 2.4 $\times10^{10}$\,erg\,cm$^{-2}$ of energy has been released. When compared to the total flare energy release, 1.2$\times10^{12}$\,erg\,cm$^{-2}$, the energy released by the start of condensation amounts to less than 2\% of the total energy. Stated differently, our model required only a small fraction of the heat flux predicted from the UFC method to produce a condensation in agreement with observation. 

By the time the condensation has sufficiently decayed, the heat supplied to the loop is an order of magnitude higher than the heat supplied during the initial, impulsive formation. This result suggests the strength of the condensation is primarily dictated by the value of flux that is initially deposited into the chromosphere, regardless of the heating profile. At the moment condensation formed in the simulation, the heating profile was releasing a flux of $F=7.5\times10^8$\,erg\,cm$^{-2}$\,s$^{-1}$. Previous numerical investigations determined a scaling relation that allowed for heat flux to be inferred from the peak downflow velocity \citep{longcope2014,ashfield2021}. Using the pre-flare coronal pressure of $p=0.7$\,erg\,cm$^{-3}$ --- the value of our model loop --- in Eq.\ (\ref{eqn:F}) yields a flux $F=7\times10^8$\,erg\,cm$^{-2}$\,s$^{-1}$. The agreement between the measured and predicted values supports the notion of the initial flux reaching the chromosphere being the primary gauge of condensation strength. Also noteworthy is how the heat flux of our model, derived using SDO and GOES coronal observations, agrees with the conclusions reached using IRIS spectroscopic measurements.

\section{Discussion} \label{sec:dis}

This work has analyzed the dynamics of a single condensation event during the X1.0 flare SOL2014-10-25T17:08:00. Using results from the previous numerical investigation in \ALpaper, we characterized the condensation decay and found the observed profile approximately follows the evolution predicted in the theoretical paper. As a consequence, the properties of the condensation were used to infer the density scale-height of the pre-flare chromosphere, where $H$=369\,km. The energy driving the condensation was inferred using the UFC method developed in \cite{qiu2012}. Incorporating footpoint emission in the AIA 1600\,\AA\,band and subsequent coronal response, the energy release for the associated flare loop was calculated to be a Gaussian heating profile with a peak flux of $F_p=6.2 \times 10^9$\,erg\,cm$^{-2}$\,s$^{-1}$ and a width of $\sigma$ = 38\,s.

To forward-model the dynamics of the observed condensation, we input the measured and inferred properties into a one-dimensional flare loop simulation run using the numerical code used in \ALpaper. The results were then used to synthesize Si \textsc{iv} 1402.77\,\AA\, spectral lines which were fit to Gaussians to find Doppler velocities which might be better compared with the IRIS SG measurements. Seen in the left panel of Figure \ref{fig:synthpreft}, the evolution of the Doppler-velocity measured from the synthetic spectra agrees reasonably well with the observed decay profile. We note, however, the brief period of increased redshifted intensity seen in observation prior to the peak downflow velocity is not reproduced by our model results. The cause of this increase was not explored in this work and remains unclear. 

We confined our analysis to a single event found by examining the evolution of the flare ribbon along the IRIS SG slit for the entire duration of the flare. Si \textsc{iv} 1402.77\,\AA\,lines were fit with an automated double-Gaussian routine in order to look for velocity profiles characteristic of condensation. A total of 60 events were found, which were subsequently divided into three types based on common features seen during their evolution. The observation of many different types of condensation is in contrast with previous studies on condensation dynamics, where flaring pixels containing condensations were shown to exhibit similar decay profiles \citep{tian2015,graham2020}. Also contrasting with other flare studies is the condensation locations within our flare ribbon. Typically, condensation is believed to occur at the leading edge of a flare ribbon, where downflows are seen to spatially and temporally coincide with new flare pixel brightenings \citep{falchi1997,graham2015,libbrecht2019}. As the ribbon evolves in this flare, however, many of the observed events occur after the initial brightening. The ribbon also remains bounded within a 4{\arcsec} region for nearly 30 minutes before diverging in two directions. Perhaps coincidentally, the persisting-type events were only found during this latter divergent phase. These events, defined by their existence within a larger decay envelope, are more consistent with condensation observations of Si \textsc{iv} that measured sustained redshifts lasting on the order of minutes \cite{warren2016,zhang2016,li2017}.

The ribbon analyzed here also developed within one of the three sunspot umbrae of AR 12192. This phenomena, known to be relatively uncommon in flares, is linked to particular characteristics, such as enhanced emission \citep{kowalski2019} and complex magnetic field structures \citep{moore1981,hofmann1987}. Indeed, the magnetic field configuration during this flare has been shown to be complicated \citep{inoue2016,bamba2017}, which may have consequences for the reconnection driving the different events. It is possible the unique aspects of the flare ribbon suggest the range of condensation behaviors seen within the ribbon are unique as well. A study characterizing condensation dynamics across a larger number of flares would likely shed light on this issue. 

Of the those identified, the pixel with the clearest condensation behavior was chosen for analysis.  This event was also found to exhibit distinctive characteristics unrelated to condensation dynamics that differentiated it from the other events in this flare. For example, the pixel of focus was part of a cluster spanning three pixels, each being classified as a archetypal-type condensation. The light curves and Doppler evolution are very similar in all three pixels, leading us to posit that a single, impulsive coronal reconnection event maps to all three.

On the other hand, the successful fit of the Si \textsc{iv} line to two Gaussian components suggests that each of the three pixels contains at least two independent plasmas.  The work presented here shows the red component evolving in a manner consistent with chromospheric condensation resulting from flare energization of the footpoint.  The velocity of the stationary component, however, does not change significantly during this time, but remains at a red shift of approximately 10\,km\,s$^{-1}$ throughout.  The simplest explanation is that the stationary component originates from a separate plasma in the pixel which is \emph{not} responding to flare energization.  To reconcile this with the three-pixel extent of the cluster, we must postulate that the coronal reconnection site maps to a distorted chromospheric region extending more than two pixels in one direction, but squashed to less than one in the other direction.  In fact, several coronal field models do predict field lines mappings around reconnection current sheets will be strongly deformed --- often by much more than 3:1 \citep{titov2002,titov2003,longcope2020}.  

A different explanation for the two-Gaussian structure was proposed by \cite{graham2020}, whose flare model was driven by a beam of non-thermal electrons. Their work suggests that with a spectrum of beam energies, electrons with the highest energies would penetrate into deep chromospheric layers, producing a strongly enhanced yet stationary emission in the line profiles. Despite the validity of this argument, we have argued against non-thermal electron deposition as the driver for the condensations observed in this work, based on the absence of HXR emission at the location of the flare ribbon and the indications of a thick-target coronal source. A different model is therefore needed to explain the observed line profiles in the event of condensations driven by thermal conduction. 

Additional support for an unresolved component is found in the longer time history of the pixel.  The same pixel hosted a weaker condensation event about 20 minutes prior to the 17:06 signature we analyzed.  If the pixel mapped to only a single reconnection site, then that site would have hosted reconnection twice.  This is not possible under conventional models of flare reconnection occurring at a single current sheet or along a single X-line  --- a pair of open field lines reconnect, become closed and remain so thereafter \citep{carmichael1964,forbes1989}.  The discussion above raises the possibility that the one pixel includes footpoints of field lines spread so far across the coronal current sheet that it was able to receive energy from two coronal reconnection events at different times. A superposition of tubes in a single pixel energized in succession would work to create a multiple-Gaussian line profile. This technique has been used in previous numerical studies studies using one-dimensional simulations to address discrepancies between models and observations \citep{reep2016,reep2018}. 

Another important consequence of this work is the connection made between chromospheric signatures and the coronal flare response. To facilitate calculation of the heat flux responsible for driving the observed condensation, we utilized a process requiring the measured coronal emission of the entire flare as an input. This was a novel application of the UFC method, which had not explored the chromospheric dynamics in response to flare heating outside the total emission in AIA\,1600\,\AA. By running our simulation with the inferred flux, we were able to effectively couple the dynamics of the lower solar atmosphere to that of the corona -- creating a holistic picture of the solar atmosphere during a flare. 

The agreement between our simulation results and the IRIS SG measurements suggest the validity of our reconnection-condensation model. This can be seen not only in the two velocity profiles, but also in the aforementioned agreement between the measured flux driving the condensation in our model and that predicted from the theoretical paper. In spite of this success, a few caveats exist pertaining to the derived heat flux. One remarkable result is the model condensation appears to depend only on the early phase of flare energy release. When energy first reaches the chromosphere, the condensation rapidly forms and begins to decay immediately thereafter, even as the energy released continues to grow in strength as dictated by our Gaussian heating profile. This result could suggest the form of the heating profile after the condensation has started is arbitrary with respect to condensation dynamics. 

Other evidence from our simulation indicates the increase in heat flux after the condensation begins to decay leads to discrepancies seen between our model and observation. One example is the decay profile, where we found the model half-life to be several seconds longer than the observed value. Although the condensation velocity in our simulation decays regardless of the continued heating, it is possible that the increase in energy deposited in the chromosphere artificially prolongs this evolution. Another example is found in the intensities of our synthesized Si \textsc{iv} spectra. After the model condensation reaches peak velocity, the intensities of the spectra continue to increase for the remainder of the condensation. This phenomena differs from the observed Si \textsc{iv} lines, which decay after peak velocity has been reached. 

The Gaussian heating profile used in this work was adopted from previous studies using the UFC method. These investigations, aiming to reproduce elongated flare emission, were able to accurately model the successive heating of flare loops assuming a Gaussian heating profile \citep{klimchuk2008,qiu2012}. However, even though the heat flux derived using the UFC method in this work produces a condensation in agreement with observation, the  explicit form of the profile is still up for debate.  In light of the discussion above, it is reasonable to assume a profile decreasing in strength after the condensation reaches a peak would produce results \revise{in better agreement} with observation. If the total energy release found from UFC is kept constant, this new profile would then require a sharper, more impulsive start. This start to flare heating is also more consistent with the energetics found in the thin flux tube version of the \textsc{PREFT} numerical code, where flare energy release occurs in the first few seconds following the retraction of a newly formed flux tube \citep{longcope2015}. Investigating the relationship of this model with the Gaussian heating profile derived from UFC, and how they pertain to chromospheric dynamics, is the aim of future work.

We finally observe the analysis presented in this paper is contingent upon a single instance of condensation emanating from a single IRIS pixel. In order for these results to be applied as a more general purpose technique for analyzing condensation behavior, more flares would need to be studied.

%\begin{acknowledgments}
%people, grant stuff..
%\end{acknowledgments}

%\vspace{5mm}
%\facilities{HST(STIS), Swift(XRT and UVOT), AAVSO, %CTIO:1.3m,
%CTIO:1.5m,CXO}

%\software{astropy %\citep{2013A&A...558A..33A,2018AJ....156..123A},  
          %Cloudy \citep{2013RMxAA..49..137F}, 
          %Source Extractor \citep{1996A&AS..117..393B}
          %}

\appendix
\section{Small and Persisting Condensation Events} 

The analysis in Section \ref{sec:spec} used a fitting routine to identify 60 condensation events over the evolution of a single flare ribbon. According to a predefined set of criteria, these events were shown to exhibit the defining characteristics of condensation, along with having significant velocities above the persistent downflow seen in Si \textsc{iv} over the lifetime of the flare ribbon. However, all but four identified condensations demonstrated atypical behavior and were subsequently classified as being either persisting- or small-type events. Here we present an example of each condensation type.

\begin{figure*}[t]
\includegraphics[width=\textwidth]{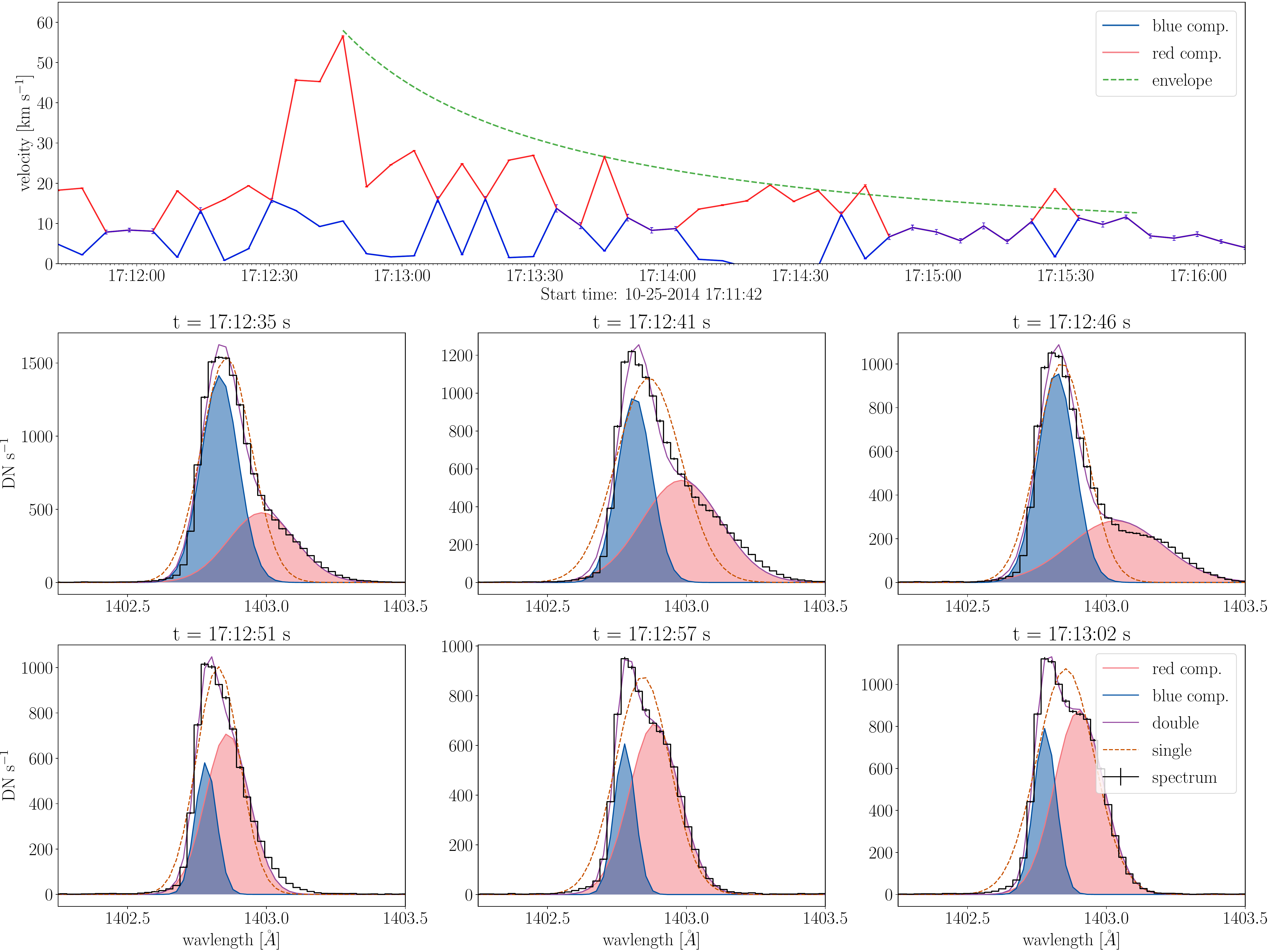}
\caption{Example evolution of a persisting-type condensation event in the same style as Figure \ref{fig:px69}. The top panel shows the velocity time series of the red and stationary components measured from the Si \textsc{iv} 1402.77\,\AA\, line profiles, along with the perceived long-term decay profile of the red-component (green dashed line).
\label{fig:app1}}
\end{figure*}

Persisting-type condensation events appeared to exhibit long-term behavior outside the typical, short-term decay profile. Figure \ref{fig:app1} shows the evolution of a single event corresponding to pixel [367.8\arcsec, -318.6\arcsec]. Shown in the top panel are the velocity time-series for the red and stationary components inferred from double-Gaussian fitting routine, respectively.
The bottom panels illustrate the line profile evolution around the time of peak downward velocity, along with the fitting results. The condensation, defined by the red-component velocity, has a slight build-up phase prior to reaching its peak at 17:12:46, after which it decays rapidly. Given our criteria, we identified this as a condensation event given this decay from nearly 60\,km\,s$^{-1}$ to 15\,km\,s$^{-1}$ in four time steps.

Notwithstanding this short-term behavior, the red-component velocity for the persisting event appears to belong to a larger decay structure. After reaching peak velocity, redshifts in the Si \textsc{iv} line decay to less than 10\,km\,s$^{-1}$ on the order of minutes. The envelope of the prolonged decay profile is shown as a green dashed line in Figure \ref{fig:app1}. All events classified as persistent exhibit this long-term behavior in addition to characteristic condensation dynamics.

\begin{figure*}[t]
\includegraphics[width=\textwidth]{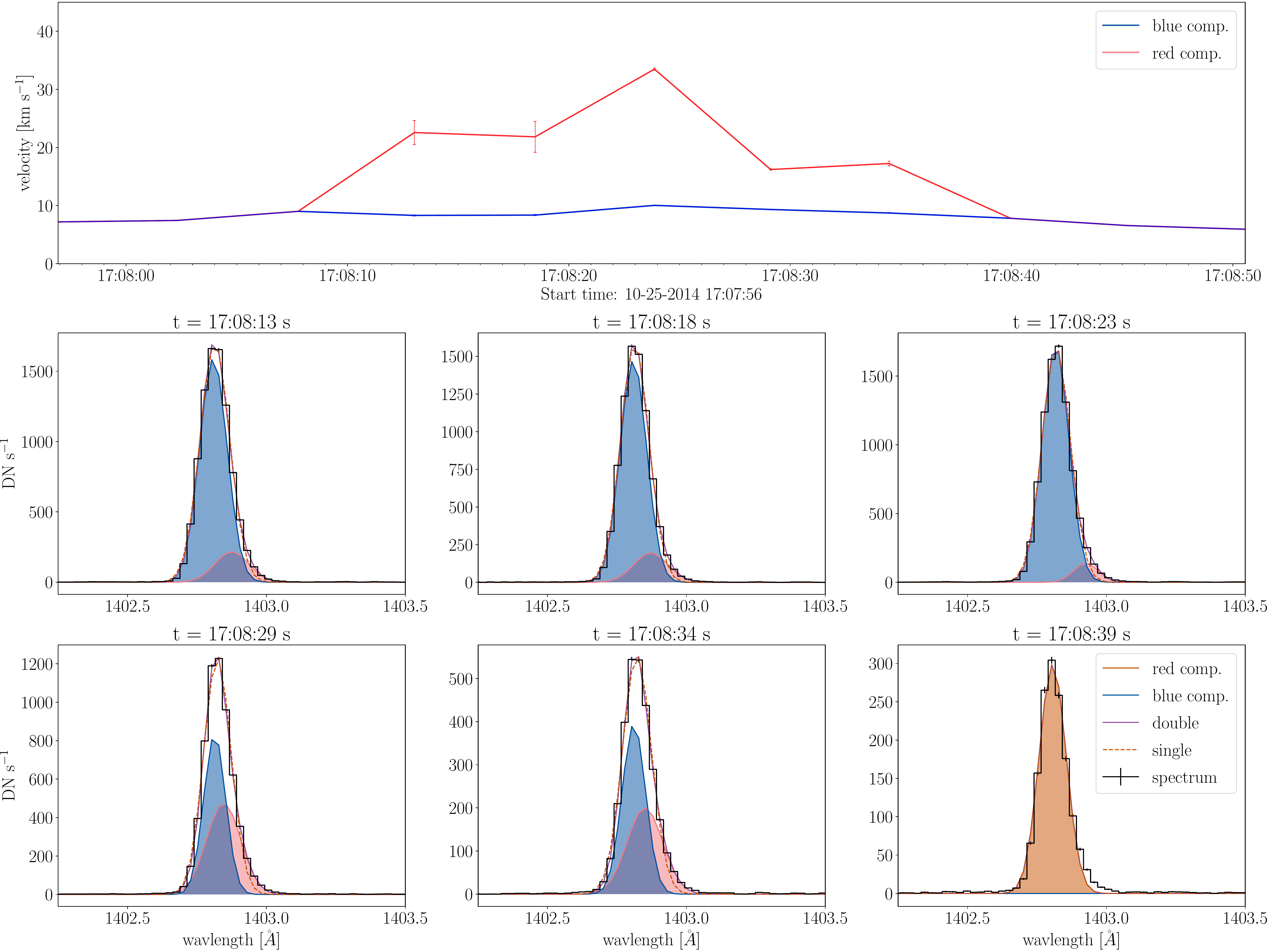}
\caption{Example evolution of a small-type condensation event in the same style as Figure \ref{fig:px69}.
\label{fig:app2}}
\end{figure*}

Small-type events have red component intensities much smaller than that of their stationary counterparts.  Figure \ref{fig:app2} shows the evolution of pixel [364.8\arcsec, -318.6\arcsec] in the same format as Figure \ref{fig:app1}. Again, there exists a build-up phase in the redshifted velocity prior to decaying back to the sustained, pre-condensation downflow. Unlike  the  line  profiles  seen  in  the  archetypal-  and persisting-type condensations, however, the profiles in small-type events characteristically lack a pronounced red asymmetry. We note that although the difference between the single- and double- Gaussian fitting results is small, the double-Gaussian fit still produces the better fit according to our reduced chi-squared threshold. We therefore must interpret the two-component Gaussian shape as an accurate representation of the line profile. Nonetheless, the intensities of the red-shifted component is small in these events, which further suggests these small-type condensations could be relatively weak.

\bibliography{short_abbrevs,master}{}

\begin{thebibliography}{}
\expandafter\ifx\csname natexlab\endcsname\relax\def\natexlab#1{#1}\fi
\providecommand{\url}[1]{\href{#1}{#1}}
\providecommand{\dodoi}[1]{doi:~\href{http://doi.org/#1}{\nolinkurl{#1}}}
\providecommand{\doeprint}[1]{\href{http://ascl.net/#1}{\nolinkurl{http://ascl.net/#1}}}
\providecommand{\doarXiv}[1]{\href{https://arxiv.org/abs/#1}{\nolinkurl{https://arxiv.org/abs/#1}}}

\bibitem[{{Acton} {et~al.}(1982){Acton}, {Leibacher}, {Canfield}, {Gunkler},
  {Hudson}, \& {Kiplinger}}]{acton1982}
{Acton}, L.~W., {Leibacher}, J.~W., {Canfield}, R.~C., {et~al.} 1982, \apj,
  263, 409, \dodoi{10.1086/160513}

\bibitem[{Allred {et~al.}(2015)Allred, Kowalski, \& Carlsson}]{allred2015}
Allred, J.~C., Kowalski, A.~F., \& Carlsson, M. 2015, The Astrophysical
  Journal, 809, 104, \dodoi{10.1088/0004-637x/809/1/104}

\bibitem[{{Antiochos} \& {Sturrock}(1978)}]{antiochos1978}
{Antiochos}, S.~K., \& {Sturrock}, P.~A. 1978, \apj, 220, 1137,
  \dodoi{10.1086/155999}

\bibitem[{{Antonucci} {et~al.}(1982){Antonucci}, {Gabriel}, {Acton}, {Culhane},
  {Doyle}, {Leibacher}, {Machado}, {Orwig}, \& {Rapley}}]{antonucci1982}
{Antonucci}, E., {Gabriel}, A.~H., {Acton}, L.~W., {et~al.} 1982, \solphys, 78,
  107, \dodoi{10.1007/BF00151147}

\bibitem[{{Ashfield} \& {Longcope}(2021)}]{ashfield2021}
{Ashfield}, W.~H., \& {Longcope}, D.~W. 2021, \apj, 912, 25,
  \dodoi{10.3847/1538-4357/abedb4}

\bibitem[{Bamba {et~al.}(2017)Bamba, Inoue, Kusano, \& Shiota}]{bamba2017}
Bamba, Y., Inoue, S., Kusano, K., \& Shiota, D. 2017, The Astrophysical
  Journal, 838, 134, \dodoi{10.3847/1538-4357/aa6682}

\bibitem[{{Battaglia} {et~al.}(2009){Battaglia}, {Fletcher}, \&
  {Benz}}]{battaglia2009}
{Battaglia}, M., {Fletcher}, L., \& {Benz}, A.~O. 2009, \aap, 498, 891,
  \dodoi{10.1051/0004-6361/200811196}

\bibitem[{{Bradshaw} \& {Cargill}(2013)}]{bradshaw2013}
{Bradshaw}, S.~J., \& {Cargill}, P.~J. 2013, \apj, 770, 12,
  \dodoi{10.1088/0004-637X/770/1/12}

\bibitem[{{Bradshaw} \& {Mason}(2003)}]{bradshaw2003}
{Bradshaw}, S.~J., \& {Mason}, H.~E. 2003, \aap, 401, 699,
  \dodoi{10.1051/0004-6361:20030089}

\bibitem[{{Brown}(1971)}]{brown1971}
{Brown}, J.~C. 1971, \solphys, 18, 489, \dodoi{10.1007/BF00149070}

\bibitem[{{Canfield} {et~al.}(1990){Canfield}, {Penn}, {Wulser}, \&
  {Kiplinger}}]{canfield1990}
{Canfield}, R.~C., {Penn}, M.~J., {Wulser}, J.-P., \& {Kiplinger}, A.~L. 1990,
  \apj, 363, 318, \dodoi{10.1086/169345}

\bibitem[{{Canfield} {et~al.}(1980){Canfield}, {Brown}, {Craig}, {Brueckner},
  {Cook}, {Doschek}, {Emslie}, {Machado}, {Henoux}, \& {Lites}}]{canfield1980}
{Canfield}, R.~C., {Brown}, J.~C., {Craig}, I.~J.~D., {et~al.} 1980, in Skylab
  Solar Workshop II, ed. P.~A. {Sturrock}, 231--271

\bibitem[{Carlsson \& Stein(1997)}]{carlsson1997}
Carlsson, M., \& Stein, R.~F. 1997, The Astrophysical Journal, 481, 500,
  \dodoi{10.1086/304043}

\bibitem[{{Carmichael}(1964)}]{carmichael1964}
{Carmichael}, H. 1964, in AAS-NASA Symposium on the Physics of Solar Flares,
  ed. W.~N. Hess (Washington, DC: NASA), 451

\bibitem[{{Chen} {et~al.}(2015){Chen}, {Zhang}, {Ma}, {Yang}, {Li}, {Huang}, \&
  {Xiao}}]{chen2015}
{Chen}, H., {Zhang}, J., {Ma}, S., {et~al.} 2015, \apjl, 808, L24,
  \dodoi{10.1088/2041-8205/808/1/L24}

\bibitem[{{Cheng} {et~al.}(1983){Cheng}, {Oran}, {Doschek}, {Boris}, \&
  {Mariska}}]{cheng1983}
{Cheng}, C.~C., {Oran}, E.~S., {Doschek}, G.~A., {Boris}, J.~P., \& {Mariska},
  J.~T. 1983, \apj, 265, 1090, \dodoi{10.1086/160751}

\bibitem[{{Coyner} \& {Alexander}(2009)}]{coyner2009}
{Coyner}, A.~J., \& {Alexander}, D. 2009, \apj, 705, 554,
  \dodoi{10.1088/0004-637X/705/1/554}

\bibitem[{{Craig} \& {McClymont}(1976)}]{craig1976}
{Craig}, I.~J.~D., \& {McClymont}, A.~N. 1976, \solphys, 50, 133,
  \dodoi{10.1007/BF00206198}

\bibitem[{{Czaykowska} {et~al.}(2001){Czaykowska}, {Alexander}, \& {De
  Pontieu}}]{czaykowska2001}
{Czaykowska}, A., {Alexander}, D., \& {De Pontieu}, B. 2001, ApJ, 552, 849,
  \dodoi{10.1086/320553}

\bibitem[{da~Costa {et~al.}(2015)da~Costa, Kleint, Petrosian, Dalda, \&
  Liu}]{rubio2015_2}
da~Costa, F.~R., Kleint, L., Petrosian, V., Dalda, A.~S., \& Liu, W. 2015, The
  Astrophysical Journal, 804, 56, \dodoi{10.1088/0004-637x/804/1/56}

\bibitem[{da~Costa {et~al.}(2016)da~Costa, Kleint, Petrosian, Liu, \&
  Allred}]{rubio2016}
da~Costa, F.~R., Kleint, L., Petrosian, V., Liu, W., \& Allred, J.~C. 2016, The
  Astrophysical Journal, 827, 38, \dodoi{10.3847/0004-637x/827/1/38}

\bibitem[{{De Pontieu} {et~al.}(2015){De Pontieu}, {McIntosh},
  {Martinez-Sykora}, {Peter}, \& {Pereira}}]{depontieu2015}
{De Pontieu}, B., {McIntosh}, S., {Martinez-Sykora}, J., {Peter}, H., \&
  {Pereira}, T.~M.~D. 2015, \apjl, 799, L12,
  \dodoi{10.1088/2041-8205/799/1/L12}

\bibitem[{{De Pontieu} {et~al.}(2014){De Pontieu}, {Title}, {Lemen}, {Kushner},
  {Akin}, {Allard}, {Berger}, {Boerner}, {Cheung}, {Chou}, {Drake}, {Duncan},
  {Freeland}, {Heyman}, {Hoffman}, {Hurlburt}, {Lindgren}, {Mathur}, {Rehse},
  {Sabolish}, {Seguin}, {Schrijver}, {Tarbell}, {W{\"u}lser}, {Wolfson},
  {Yanari}, {Mudge}, {Nguyen-Phuc}, {Timmons}, {van Bezooijen}, {Weingrod},
  {Brookner}, {Butcher}, {Dougherty}, {Eder}, {Knagenhjelm}, {Larsen},
  {Mansir}, {Phan}, {Boyle}, {Cheimets}, {DeLuca}, {Golub}, {Gates}, {Hertz},
  {McKillop}, {Park}, {Perry}, {Podgorski}, {Reeves}, {Saar}, {Testa}, {Tian},
  {Weber}, {Dunn}, {Eccles}, {Jaeggli}, {Kankelborg}, {Mashburn}, {Pust},
  {Springer}, {Carvalho}, {Kleint}, {Marmie}, {Mazmanian}, {Pereira}, {Sawyer},
  {Strong}, {Worden}, {Carlsson}, {Hansteen}, {Leenaarts}, {Wiesmann},
  {Aloise}, {Chu}, {Bush}, {Scherrer}, {Brekke}, {Martinez-Sykora}, {Lites},
  {McIntosh}, {Uitenbroek}, {Okamoto}, {Gummin}, {Auker}, {Jerram}, {Pool}, \&
  {Waltham}}]{depontieu2014}
{De Pontieu}, B., {Title}, A.~M., {Lemen}, J.~R., {et~al.} 2014, \solphys, 289,
  2733, \dodoi{10.1007/s11207-014-0485-y}

\bibitem[{De~Pontieu {et~al.}(2021)De~Pontieu, Polito, Hansteen, Testa, Reeves,
  Antolin, N{\'o}brega-Siverio, Kowalski, Martinez-Sykora, Carlsson, McIntosh,
  Liu, Daw, \& Kankelborg}]{depontieu2021}
De~Pontieu, B., Polito, V., Hansteen, V., {et~al.} 2021, Solar Physics, 296,
  84, \dodoi{10.1007/s11207-021-01826-0}

\bibitem[{{Del Zanna} {et~al.}(2015){Del Zanna}, {Dere}, {Young}, {Landi}, \&
  {Mason}}]{chianti}
{Del Zanna}, G., {Dere}, K.~P., {Young}, P.~R., {Landi}, E., \& {Mason}, H.~E.
  2015, AAP, 582, A56, \dodoi{10.1051/0004-6361/201526827}

\bibitem[{{Doschek} {et~al.}(1976){Doschek}, {Feldman}, \&
  {Bohlin}}]{doschek1976}
{Doschek}, G.~A., {Feldman}, U., \& {Bohlin}, J.~D. 1976, \apjl, 205, L177,
  \dodoi{10.1086/182118}

\bibitem[{Dud{\'\i}k {et~al.}(2017)Dud{\'\i}k, Dzif{\v c}{\'a}kov{\'a},
  Meyer-Vernet, Del~Zanna, Young, Giunta, Sylwester, Sylwester, Oka, Mason,
  Vocks, Matteini, Krucker, Williams, \& Mackovjak}]{dudik2017}
Dud{\'\i}k, J., Dzif{\v c}{\'a}kov{\'a}, E., Meyer-Vernet, N., {et~al.} 2017,
  Solar Physics, 292, 100, \dodoi{10.1007/s11207-017-1125-0}

\bibitem[{{Emslie} \& {Sturrock}(1982)}]{emslie1982}
{Emslie}, A.~G., \& {Sturrock}, P.~A. 1982, \solphys, 80, 99,
  \dodoi{10.1007/BF00153426}

\bibitem[{{Falchi} {et~al.}(1997){Falchi}, {Qiu}, \& {Cauzzi}}]{falchi1997}
{Falchi}, A., {Qiu}, J., \& {Cauzzi}, G. 1997, \aap, 328, 371

\bibitem[{{Fisher}(1986)}]{fisher1986_2}
{Fisher}, G.~H. 1986, in The Lower Atmosphere of Solar Flares, ed. D.~F.
  {Neidig} \& M.~E. {Machado}, 25--36

\bibitem[{Fisher(1987)}]{fisher1987}
Fisher, G.~H. 1987, Solar Physics, 113, 307, \dodoi{10.1007/BF00147715}

\bibitem[{Fisher(1989)}]{fisher1989}
---. 1989, APJ, 346

\bibitem[{{Fisher} {et~al.}(1985{\natexlab{a}}){Fisher}, {Canfield}, \&
  {McClymont}}]{fisher1985_1}
{Fisher}, G.~H., {Canfield}, R.~C., \& {McClymont}, A.~N. 1985{\natexlab{a}},
  \apj, 289, 425, \dodoi{10.1086/162902}

\bibitem[{{Fisher} {et~al.}(1985{\natexlab{b}}){Fisher}, {Canfield}, \&
  {McClymont}}]{fisher1985_2}
---. 1985{\natexlab{b}}, \apj, 289, 434, \dodoi{10.1086/162903}

\bibitem[{{Fletcher} \& {Hudson}(2008)}]{fletcher2008}
{Fletcher}, L., \& {Hudson}, H.~S. 2008, \apj, 675, 1645,
  \dodoi{10.1086/527044}

\bibitem[{{Fletcher} {et~al.}(2011){Fletcher}, {Dennis}, {Hudson}, {Krucker},
  {Phillips}, {Veronig}, {Battaglia}, {Bone}, {Caspi}, {Chen}, {Gallagher},
  {Grigis}, {Ji}, {Liu}, {Milligan}, \& {Temmer}}]{fletcher2011}
{Fletcher}, L., {Dennis}, B.~R., {Hudson}, H.~S., {et~al.} 2011, \ssr, 159, 19,
  \dodoi{10.1007/s11214-010-9701-8}

\bibitem[{{Forbes} {et~al.}(1989){Forbes}, {Malherbe}, \&
  {Priest}}]{forbes1989}
{Forbes}, T.~G., {Malherbe}, J.~M., \& {Priest}, E.~R. 1989, Solar~Phys., 120,
  285, \dodoi{10.1007/BF00159881}

\bibitem[{{Freeland} \& {Handy}(1998)}]{freeland1998}
{Freeland}, S.~L., \& {Handy}, B.~N. 1998, \solphys, 182, 497,
  \dodoi{10.1023/A:1005038224881}

\bibitem[{{Gan} {et~al.}(1991){Gan}, {Zhang}, \& {Fang}}]{gan1991}
{Gan}, W.~Q., {Zhang}, H.~Q., \& {Fang}, C. 1991, \aap, 241, 618

\bibitem[{{Graham} \& {Cauzzi}(2015)}]{graham2015}
{Graham}, D.~R., \& {Cauzzi}, G. 2015, \apjl, 807, L22,
  \dodoi{10.1088/2041-8205/807/2/L22}

\bibitem[{{Graham} {et~al.}(2020){Graham}, {Cauzzi}, {Zangrilli}, {Kowalski},
  {Sim{\~o}es}, \& {Allred}}]{graham2020}
{Graham}, D.~R., {Cauzzi}, G., {Zangrilli}, L., {et~al.} 2020, \apj, 895, 6,
  \dodoi{10.3847/1538-4357/ab88ad}

\bibitem[{{Guidoni} \& {Longcope}(2010)}]{guidoni2010}
{Guidoni}, S.~E., \& {Longcope}, D.~W. 2010, \apj, 718, 1476,
  \dodoi{10.1088/0004-637X/718/2/1476}

\bibitem[{{Guo} {et~al.}(2012){Guo}, {Emslie}, {Massone}, \&
  {Piana}}]{guo2012_1}
{Guo}, J., {Emslie}, A.~G., {Massone}, A.~M., \& {Piana}, M. 2012, \apj, 755,
  32, \dodoi{10.1088/0004-637X/755/1/32}

\bibitem[{{Hansteen}(1993)}]{hansteen1993}
{Hansteen}, V. 1993, \apj, 402, 741, \dodoi{10.1086/172174}

\bibitem[{{Hofmann} {et~al.}(1987){Hofmann}, {Rendtel}, {Aurass}, \&
  {Kalman}}]{hofmann1987}
{Hofmann}, A., {Rendtel}, J., {Aurass}, H., \& {Kalman}, B. 1987, \solphys,
  108, 151, \dodoi{10.1007/BF00152084}

\bibitem[{{Holman} {et~al.}(2011){Holman}, {Aschwanden}, {Aurass}, {Battaglia},
  {Grigis}, {Kontar}, {Liu}, {Saint-Hilaire}, \& {Zharkova}}]{holman2011}
{Holman}, G.~D., {Aschwanden}, M.~J., {Aurass}, H., {et~al.} 2011, \ssr, 159,
  107, \dodoi{10.1007/s11214-010-9680-9}

\bibitem[{{Inoue} {et~al.}(2016){Inoue}, {Hayashi}, \& {Kusano}}]{inoue2016}
{Inoue}, S., {Hayashi}, K., \& {Kusano}, K. 2016, \apj, 818, 168,
  \dodoi{10.3847/0004-637X/818/2/168}

\bibitem[{{Kerr} {et~al.}(2019){Kerr}, {Carlsson}, \& {Allred}}]{kerr2019s}
{Kerr}, G.~S., {Carlsson}, M., \& {Allred}, J.~C. 2019, \apj, 885, 119,
  \dodoi{10.3847/1538-4357/ab48ea}

\bibitem[{Kleint {et~al.}(2017)Kleint, Heinzel, \& Krucker}]{kleint2017}
Kleint, L., Heinzel, P., \& Krucker, S. 2017, The Astrophysical Journal, 837,
  160, \dodoi{10.3847/1538-4357/aa62fe}

\bibitem[{{Klimchuk} {et~al.}(2008){Klimchuk}, {Patsourakos}, \&
  {Cargill}}]{klimchuk2008}
{Klimchuk}, J.~A., {Patsourakos}, S., \& {Cargill}, P.~J. 2008, \apj, 682,
  1351, \dodoi{10.1086/589426}

\bibitem[{{Kontar} {et~al.}(2011){Kontar}, {Brown}, {Emslie}, {Hajdas},
  {Holman}, {Hurford}, {Ka{\v{s}}parov{\'a}}, {Mallik}, {Massone}, {McConnell},
  {Piana}, {Prato}, {Schmahl}, \& {Suarez-Garcia}}]{kontar2011}
{Kontar}, E.~P., {Brown}, J.~C., {Emslie}, A.~G., {et~al.} 2011, \ssr, 159,
  301, \dodoi{10.1007/s11214-011-9804-x}

\bibitem[{{Kopp} \& {Pneuman}(1976)}]{kopp1976}
{Kopp}, R.~A., \& {Pneuman}, G.~W. 1976, \solphys, 50, 85,
  \dodoi{10.1007/BF00206193}

\bibitem[{{Kowalski} {et~al.}(2017){Kowalski}, {Allred}, {Daw}, {Cauzzi}, \&
  {Carlsson}}]{kowalski2017}
{Kowalski}, A.~F., {Allred}, J.~C., {Daw}, A., {Cauzzi}, G., \& {Carlsson}, M.
  2017, \apj, 836, 12, \dodoi{10.3847/1538-4357/836/1/12}

\bibitem[{Kowalski {et~al.}(2019)Kowalski, Butler, Daw, Fletcher, Allred,
  Pontieu, Kerr, \& Cauzzi}]{kowalski2019}
Kowalski, A.~F., Butler, E., Daw, A.~N., {et~al.} 2019, The Astrophysical
  Journal, 878, 135, \dodoi{10.3847/1538-4357/ab1f8b}

\bibitem[{Krucker {et~al.}(2008)Krucker, Battaglia, Cargill, Fletcher, Hudson,
  MacKinnon, Masuda, Sui, Tomczak, Veronig, Vlahos, \& White}]{krucker2008}
Krucker, S., Battaglia, M., Cargill, P.~J., {et~al.} 2008, The Astronomy and
  Astrophysics Review, 16, 155, \dodoi{10.1007/s00159-008-0014-9}

\bibitem[{Kuridze {et~al.}(2015)Kuridze, Mathioudakis, Sim{\~{o}}es, van~der
  Voort, Carlsson, Jafarzadeh, Allred, Kowalski, Kennedy, Fletcher, Graham, \&
  Keenan}]{kuridze2015}
Kuridze, D., Mathioudakis, M., Sim{\~{o}}es, P. J.~A., {et~al.} 2015, The
  Astrophysical Journal, 813, 125, \dodoi{10.1088/0004-637x/813/2/125}

\bibitem[{{Lemen} {et~al.}(2012){Lemen}, {Title}, {Akin}, {Boerner}, {Chou},
  {Drake}, {Duncan}, {Edwards}, {Friedlaender}, {Heyman}, {Hurlburt}, {Katz},
  {Kushner}, {Levay}, {Lindgren}, {Mathur}, {McFeaters}, {Mitchell}, {Rehse},
  {Schrijver}, {Springer}, {Stern}, {Tarbell}, {Wuelser}, {Wolfson}, {Yanari},
  {Bookbinder}, {Cheimets}, {Caldwell}, {Deluca}, {Gates}, {Golub}, {Park},
  {Podgorski}, {Bush}, {Scherrer}, {Gummin}, {Smith}, {Auker}, {Jerram},
  {Pool}, {Soufli}, {Windt}, {Beardsley}, {Clapp}, {Lang}, \&
  {Waltham}}]{lemen2012}
{Lemen}, J.~R., {Title}, A.~M., {Akin}, D.~J., {et~al.} 2012, Solar~Phys., 275,
  17, \dodoi{10.1007/s11207-011-9776-8}

\bibitem[{Li {et~al.}(2018)Li, Zhang, Yang, \& Hou}]{li2019_cc}
Li, X., Zhang, J., Yang, S., \& Hou, Y. 2018, Publications of the Astronomical
  Society of Japan, 71, \dodoi{10.1093/pasj/psy128}

\bibitem[{{Li} {et~al.}(2017){Li}, {Kelly}, {Ding}, {Qiu}, {Zhu}, \&
  {Gan}}]{li2017}
{Li}, Y., {Kelly}, M., {Ding}, M.~D., {et~al.} 2017, apj, 848, 118,
  \dodoi{10.3847/1538-4357/aa89e4}

\bibitem[{{Libbrecht} {et~al.}(2019){Libbrecht}, {de la Cruz Rodr{\'\i}guez},
  {Danilovic}, {Leenaarts}, \& {Pazira}}]{libbrecht2019}
{Libbrecht}, T., {de la Cruz Rodr{\'\i}guez}, J., {Danilovic}, S., {Leenaarts},
  J., \& {Pazira}, H. 2019, \aap, 621, A35, \dodoi{10.1051/0004-6361/201833610}

\bibitem[{{Lin} {et~al.}(2002){Lin}, {Dennis}, {Hurford}, {Smith}, {Zehnder},
  {Harvey}, {Curtis}, {Pankow}, {Turin}, {Bester}, {Csillaghy}, {Lewis},
  {Madden}, {van Beek}, {Appleby}, {Raudorf}, {McTiernan}, {Ramaty}, {Schmahl},
  {Schwartz}, {Krucker}, {Abiad}, {Quinn}, {Berg}, {Hashii}, {Sterling},
  {Jackson}, {Pratt}, {Campbell}, {Malone}, {Landis}, {Barrington-Leigh},
  {Slassi-Sennou}, {Cork}, {Clark}, {Amato}, {Orwig}, {Boyle}, {Banks},
  {Shirey}, {Tolbert}, {Zarro}, {Snow}, {Thomsen}, {Henneck}, {McHedlishvili},
  {Ming}, {Fivian}, {Jordan}, {Wanner}, {Crubb}, {Preble}, {Matranga}, {Benz},
  {Hudson}, {Canfield}, {Holman}, {Crannell}, {Kosugi}, {Emslie}, {Vilmer},
  {Brown}, {Johns-Krull}, {Aschwanden}, {Metcalf}, \& {Conway}}]{lin2002}
{Lin}, R.~P., {Dennis}, B.~R., {Hurford}, G.~J., {et~al.} 2002, \solphys, 210,
  3, \dodoi{10.1023/A:1022428818870}

\bibitem[{Liu {et~al.}(2013)Liu, Qiu, Longcope, \& Caspi}]{liu2013}
Liu, W.-J., Qiu, J., Longcope, D.~W., \& Caspi, A. 2013, The Astrophysical
  Journal, 770, 111, \dodoi{10.1088/0004-637x/770/2/111}

\bibitem[{Longcope {et~al.}(2020)Longcope, McCarthy, \&
  Malanushenko}]{longcope2020}
Longcope, D., McCarthy, M., \& Malanushenko, A. 2020, The Astrophysical
  Journal, 901, 147, \dodoi{10.3847/1538-4357/abb2a9}

\bibitem[{Longcope(2014)}]{longcope2014}
Longcope, D.~W. 2014, APJ, 795

\bibitem[{{Longcope} {et~al.}(2009){Longcope}, {Guidoni}, \&
  {Linton}}]{longcope2009}
{Longcope}, D.~W., {Guidoni}, S.~E., \& {Linton}, M.~G. 2009, \apjl, 690, L18,
  \dodoi{10.1088/0004-637X/690/1/L18}

\bibitem[{Longcope \& Klimchuk(2015)}]{longcope2015}
Longcope, D.~W., \& Klimchuk, J.~A. 2015, The Astrophysical Journal, 813, 131,
  \dodoi{10.1088/0004-637x/813/2/131}

\bibitem[{{MacNeice}(1986)}]{macneice1986}
{MacNeice}, P. 1986, \solphys, 103, 47, \dodoi{10.1007/BF00154858}

\bibitem[{{Moore}(1981)}]{moore1981}
{Moore}, R.~L. 1981, \ssr, 28, 387, \dodoi{10.1007/BF00212601}

\bibitem[{{Nagai} \& {Emslie}(1984)}]{nagai1984}
{Nagai}, F., \& {Emslie}, A.~G. 1984, \apj, 279, 896, \dodoi{10.1086/161960}

\bibitem[{{Neupert}(1968)}]{neupert1968}
{Neupert}, W.~M. 1968, \apjl, 153, L59, \dodoi{10.1086/180220}

\bibitem[{Priest \& Forbes(2002)}]{priest2002}
Priest, E.~R., \& Forbes, T.~G. 2002, The Astronomy and Astrophysics Review,
  10, 313, \dodoi{10.1007/s001590100013}

\bibitem[{Qiu(2021)}]{qiu2021}
Qiu, J. 2021, The Astrophysical Journal, 909, 99,
  \dodoi{10.3847/1538-4357/abe0b3}

\bibitem[{Qiu {et~al.}(2012)Qiu, Liu, \& Longcope}]{qiu2012}
Qiu, J., Liu, W.-J., \& Longcope, D.~W. 2012, The Astrophysical Journal, 752,
  124, \dodoi{10.1088/0004-637x/752/2/124}

\bibitem[{Qiu \& Longcope(2016)}]{qiu2016}
Qiu, J., \& Longcope, D.~W. 2016, The Astrophysical Journal, 820, 14,
  \dodoi{10.3847/0004-637x/820/1/14}

\bibitem[{{Reep} {et~al.}(2018){Reep}, {Polito}, {Warren}, \&
  {Crump}}]{reep2018}
{Reep}, J.~W., {Polito}, V., {Warren}, H.~P., \& {Crump}, N.~A. 2018, \apj,
  856, 149, \dodoi{10.3847/1538-4357/aab273}

\bibitem[{{Reep} {et~al.}(2016){Reep}, {Warren}, {Crump}, \&
  {Sim{\~o}es}}]{reep2016}
{Reep}, J.~W., {Warren}, H.~P., {Crump}, N.~A., \& {Sim{\~o}es}, P. J.~A. 2016,
  \apj, 827, 145, \dodoi{10.3847/0004-637X/827/2/145}

\bibitem[{Rosner {et~al.}(1978)Rosner, Tucker, \& Vaiana}]{rosner1978}
Rosner, R., Tucker, W.~H., \& Vaiana, G.~S. 1978, ApJ, 220, 643

\bibitem[{{Rubio da Costa} {et~al.}(2015){Rubio da Costa}, {Liu}, {Petrosian},
  \& {Carlsson}}]{rubio2015}
{Rubio da Costa}, F., {Liu}, W., {Petrosian}, V., \& {Carlsson}, M. 2015, \apj,
  813, 133, \dodoi{10.1088/0004-637X/813/2/133}

\bibitem[{Rybicki \& Lightman(1985)}]{rybicki1985}
Rybicki, G.~B., \& Lightman, A.~P. 1985, RADIATIVE TRANSITIONS (John Wiley \&
  Sons, Ltd), 267--293, \dodoi{https://doi.org/10.1002/9783527618170.ch10}

\bibitem[{{Scherrer} {et~al.}(2012){Scherrer}, {Schou}, {Bush}, {Kosovichev},
  {Bogart}, {Hoeksema}, {Liu}, {Duvall}, {Zhao}, {Title}, {Schrijver},
  {Tarbell}, \& {Tomczyk}}]{schou2012}
{Scherrer}, P.~H., {Schou}, J., {Bush}, R.~I., {et~al.} 2012, \solphys, 275,
  207, \dodoi{10.1007/s11207-011-9834-2}

\bibitem[{{Shapiro} \& {Moore}(1977)}]{shapiro1977}
{Shapiro}, P.~R., \& {Moore}, R.~T. 1977, \apj, 217, 621,
  \dodoi{10.1086/155609}

\bibitem[{{Shen} {et~al.}(2013){Shen}, {Reeves}, {Raymond}, {Murphy}, {Ko},
  {Lin}, {Miki{\'c}}, \& {Linker}}]{shen2013}
{Shen}, C., {Reeves}, K.~K., {Raymond}, J.~C., {et~al.} 2013, \apj, 773, 110,
  \dodoi{10.1088/0004-637X/773/2/110}

\bibitem[{Sui {et~al.}(2004)Sui, Holman, \& Dennis}]{sui2004}
Sui, L., Holman, G.~D., \& Dennis, B.~R. 2004, The Astrophysical Journal, 612,
  546, \dodoi{10.1086/422515}

\bibitem[{{Syrovatskii} \& {Shmeleva}(1972)}]{syrovatskii1972}
{Syrovatskii}, S.~I., \& {Shmeleva}, O.~P. 1972, \sovast, 16, 273

\bibitem[{{Tian} {et~al.}(2015){Tian}, {Young}, {Reeves}, {Chen}, {Liu}, \&
  {McKillop}}]{tian2015}
{Tian}, H., {Young}, P.~R., {Reeves}, K.~K., {et~al.} 2015, apj, 811, 139,
  \dodoi{10.1088/0004-637X/811/2/139}

\bibitem[{Titov {et~al.}(2003)Titov, Galsgaard, \& Neukirch}]{titov2003}
Titov, V.~S., Galsgaard, K., \& Neukirch, T. 2003, The Astrophysical Journal,
  582, 1172, \dodoi{10.1086/344799}

\bibitem[{Titov {et~al.}(2002)Titov, Hornig, \&
  D{\~A}{\copyright}moulin}]{titov2002}
Titov, V.~S., Hornig, G., \& D{\~A}{\copyright}moulin, P. 2002, Journal of
  Geophysical Research: Space Physics, 107, SSH 3,
  \dodoi{https://doi.org/10.1029/2001JA000278}

\bibitem[{{Veronig} {et~al.}(2002){Veronig}, {Vr{\v{s}}nak}, {Dennis},
  {Temmer}, {Hanslmeier}, \& {Magdaleni{\'c}}}]{veronig2002}
{Veronig}, A., {Vr{\v{s}}nak}, B., {Dennis}, B.~R., {et~al.} 2002, \aap, 392,
  699, \dodoi{10.1051/0004-6361:20020947}

\bibitem[{Veronig \& Brown(2004)}]{veronig2004}
Veronig, A.~M., \& Brown, J.~C. 2004, The Astrophysical Journal, 603, L117,
  \dodoi{10.1086/383199}

\bibitem[{Warren {et~al.}(2016)Warren, Reep, Crump, \&
  Sim{\~{o}}es}]{warren2016}
Warren, H.~P., Reep, J.~W., Crump, N.~A., \& Sim{\~{o}}es, P. J.~A. 2016, The
  Astrophysical Journal, 829, 35, \dodoi{10.3847/0004-637x/829/1/35}

\bibitem[{{Warren} \& {Warshall}(2001)}]{warren2001}
{Warren}, H.~P., \& {Warshall}, A.~D. 2001, \apjl, 560, L87,
  \dodoi{10.1086/324060}

\bibitem[{{Zarro} \& {Canfield}(1989)}]{zarro1989}
{Zarro}, D.~M., \& {Canfield}, R.~C. 1989, \apjl, 338, L33,
  \dodoi{10.1086/185394}

\bibitem[{{Zarro} \& {Lemen}(1988)}]{zarro1988}
{Zarro}, D.~M., \& {Lemen}, J.~R. 1988, ApJ, 329, 456, \dodoi{10.1086/166391}

\bibitem[{{Zhang} {et~al.}(2016){Zhang}, {Li}, \& {Ning}}]{zhang2016}
{Zhang}, Q.~M., {Li}, D., \& {Ning}, Z.~J. 2016, \apj, 832, 65,
  \dodoi{10.3847/0004-637X/832/1/65}

\bibitem[{Zhu {et~al.}(2018)Zhu, Qiu, \& Longcope}]{zhu2018}
Zhu, C., Qiu, J., \& Longcope, D.~W. 2018, The Astrophysical Journal, 856, 27,
  \dodoi{10.3847/1538-4357/aaad10}

\end{thebibliography}
\bibliographystyle{aasjournal}

\end{document}